\documentclass{article}

\PassOptionsToPackage{numbers,compress}{natbib}
\usepackage[preprint]{neurips_2025}
\usepackage{amsmath}

\usepackage{newtxtext,newtxmath}
\usepackage{threeparttable}

\usepackage{booktabs}
\usepackage{multirow}

\usepackage{graphicx}
\usepackage{url}
\usepackage{hyperref}
\hypersetup{colorlinks=true, linkcolor=blue, citecolor=blue, urlcolor=blue}

\makeatletter
\renewcommand{\fnum@figure}{\textbf{Figure \thefigure}}
\renewcommand{\fnum@table}{\textbf{Table \thetable}}
\makeatother

\def\papertitle{
Missing the Butterfly and Predicting the Past: Features or Bugs of Accurate AI Weather Models?
}

\title{\papertitle}

\author{%
Pedram Hassanzadeh$^{1,2,3\dagger\ast}$
\And Weidong Li$^{1\dagger}$
\And Y. Qiang Sun$^{4,5,1\ast}$
\And Jiangdi Wang$^{2}$
\And Alexander Wikner$^{1,3}$
\And Justin Finkel$^{3}$
\And Jonathan Q. Weare$^{6}$
\AND
{\normalfont \small $^{1}$Department of Geophysical Sciences, University of Chicago, Chicago, USA} \\
{\normalfont \small $^{2}$Committee on Computational and Applied Mathematics, University of Chicago, Chicago, USA.} \\
{\normalfont \small $^{3}$Data Science Institute, University of Chicago, Chicago, USA.} \\ 
{\normalfont \small $^{4}$State Key Laboratory of Severe Weather Meteorological Science and Technology, Nanjing University, Nanjing, China.}   \\
{\normalfont \small $^{5}$Key Laboratory of Mesoscale Severe Weather/Ministry of Education, Nanjing University, Nanjing, China.}   \\
{\normalfont \small $^{6}$Courant Institute School of Mathematics, Computing, and Data Science, New York University, New York City, USA.}  \\
\AND
{\normalfont \small $^\ast$Corresponding authors: pedramh@uchicago.edu, qiangsun@nju.edu.cn} \\
\normalfont {\small $^\dagger$These authors contributed equally to this work.}
}

\begin{document} 

\maketitle

\begin{abstract} 
AI weather prediction (AIWP) models rival physics-based models, yet the sources of their unexpected forecast accuracy and the degree of their physical fidelity remain unclear. Here, across a hierarchy spanning observation-based reanalysis, a general circulation model, and the multi-scale Lorenz system, we show that AI models can be trained to skillfully predict the past (backcast), though backcasts are systematically less accurate than forecasts. However, skillful backcasting appears to violate the second law of thermodynamics, and all these forecasting and backcasting models miss the butterfly effect. We trace the surprising forecast accuracy, missing butterfly, and skillful backcasting to a single cause: inevitable coarse-graining of training data, which removes fast, small scales and/or some variables. From the Lorenz system to official Pangu-Weather models, reducing coarse-graining makes AI predictions more physics-like (arrow of time and butterfly-like effects emerge), but forecast accuracy declines. Results offer an explanation for AIWP models' forecast skill: unlike physics-based models, they implicitly learn how fast, small scales affect large scales without inheriting their rapid error growth. Broader implications are that AI models' proliferation calls for revisiting predictability theories and long-term climate emulation strategies, and backcasting offers a useful, new lens for such analyses.

\end{abstract}


\section{Introduction}
Artificial intelligence (AI) weather prediction (AIWP) models have transformed short- to medium-range (days to two weeks) forecasting \cite{pathak2022fourcastnet,bi2023accurate,lam2023learning}: they outperform the best physics-based numerical weather prediction (NWP) models across a wide range of common skill metrics at orders of magnitude lower computational cost \cite{price2025probabilistic, kochkov2024neural, rasp2024weatherbench, masiwal2026decision}. State-of-the-art AIWP models are trained in a broadly similar fashion: to advance the atmosphere's 3D global state $\mathbf{x}$ from time $t$ to $t+\Delta t$ using a deep neural network $\mathcal{NN}$, i.e., 
\begin{equation}
  \mathbf{x}(t+\Delta t)=\mathcal{NN}_ \theta\big(\mathbf{x}(t)\big), \label{eq:FC}
\end{equation}
where $\Delta t$ is often 6-24 hours, $\theta$ represents the learnable parameters, and the training data come from $\sim 40$ years of observation-based reanalysis products like ERA5 \cite{hersbach2020era5}. Longer forecasts are produced autoregressively. Given their accuracy and computational efficiency, AIWP models are now being operationalized by leading forecasting centers \cite{daub2025fastnet,moldovan2026aifs} and increasingly used to provide actionable weather information to those most in need \cite{aitken2026designing}. More recently, these models have been pushed to subseasonal-to-seasonal time scales and even to longer-term climate projections, with promising early results~\cite{watt2024ace2,henn2026aimip}.

Yet, a growing number of studies have identified several \emph{failure modes} of AIWP models, e.g., their inability to forecast unprecedented ``gray swan'' weather extremes~\cite{sun2025can,sun2025predicting,zhang2026physics} and to reproduce the small-scales' statistics~\cite{chattopadhyay2023challenges, bonavita2024some}. Most surprisingly, these highly skillful models were found to lack the \emph{butterfly effect} \cite{selz2023can}: the rapid growth of ensemble spread, which accelerates as initial-condition perturbations vanish, in multi-scale chaotic systems~\cite{lorenz1969predictability, palmer2014real, palmer2024real}.
This deficiency appears shared across otherwise very different deterministic and generative AI architectures~\cite{selz2026can}. {\it While this failure has generated concerns, its root cause and implications remain unclear}. Meanwhile, the sources of AIWP models' surprising accuracy are also poorly 
understood, hindering trust and further improvements. Indeed, this accuracy comes despite long-standing skepticism about skillful data-driven weather forecasting, e.g., from Lorenz~\cite{lorenz1969atmospheric} and 
others, including an estimated requirement of $\sim\!10^{30}$ years of training data~\cite{van1994searching}; see Discussion. Underlying these puzzles, fundamental questions about the chaotic atmosphere's predictability limits, and the butterfly effect's practical relevance, remain a subject of extensive research and debate \cite{lorenz1996predictability, durran2014atmospheric,  sun2016intrinsic, zhang2019predictability, vonich2026atmospheric}.

Here, we investigate the missing butterfly effect and surprising forecast accuracy of AIWP models. First, since nothing in these AIWP models' design restricts them to the arrow of time, we also train ``backcasting'' models: neural networks that predict the past (Fig.~\ref{fig1:era5-plasim}A), i.e.,
\begin{equation}
  \mathbf{x}(t-\Delta t)=\mathcal{NN}_\phi\big(\mathbf{x}(t) \big), \label{eq:BC}
\end{equation}
the Eq.~\eqref{eq:FC} ``forecast'' model's time-reversed counterpart, with its own (independent) learnable parameters $\phi$. As shown below, autoregressive backcasting offers a new, useful dimension for exploring AIWP models and fundamentals of atmospheric predictability. 

Across a hierarchy spanning ERA5 reanalysis, an intermediate-complexity general circulation model (GCM), and multi-scale Lorenz 96, we show that, while skillful, these backcasting AI models consistently behave in ways sharply at odds with fundamental understanding of dynamical systems and their numerical integrations. Simply put, skillful backcasting appears to violate the second law of thermodynamics. {\it We also show that across the hierarchy, the butterfly effect is missing from these AI forecasting and backcasting models alike}. We then explicitly connect all three (surprising AI forecast skills, the missing butterfly effect, skillful AI backcasting), showing they are symptoms of the same underlying cause. We examine whether each symptom is a feature or a bug of skillful AIWP models.

\section{Results}
For each system in our hierarchy, the backcasting AI models (Eq.~\eqref{eq:BC}) are architecturally identical to the forecasting models (Eq.~\eqref{eq:FC}) and are trained on the same data, optimizer, and loss function (often the mean absolute or squared error, MAE/MSE); the sole difference is the reversed input--output pairing  (Fig.~\ref{fig1:era5-plasim}A). We have trained a number of deterministic and generative AIWP models based on state-of-the-art architectures (see Methods and Data).

\subsection{Skillful AI backcasting and persistent asymmetry: ERA5 and PlaSim (GCM)}
\begin{figure}[t]
	\centering
	\includegraphics[width=1\textwidth]{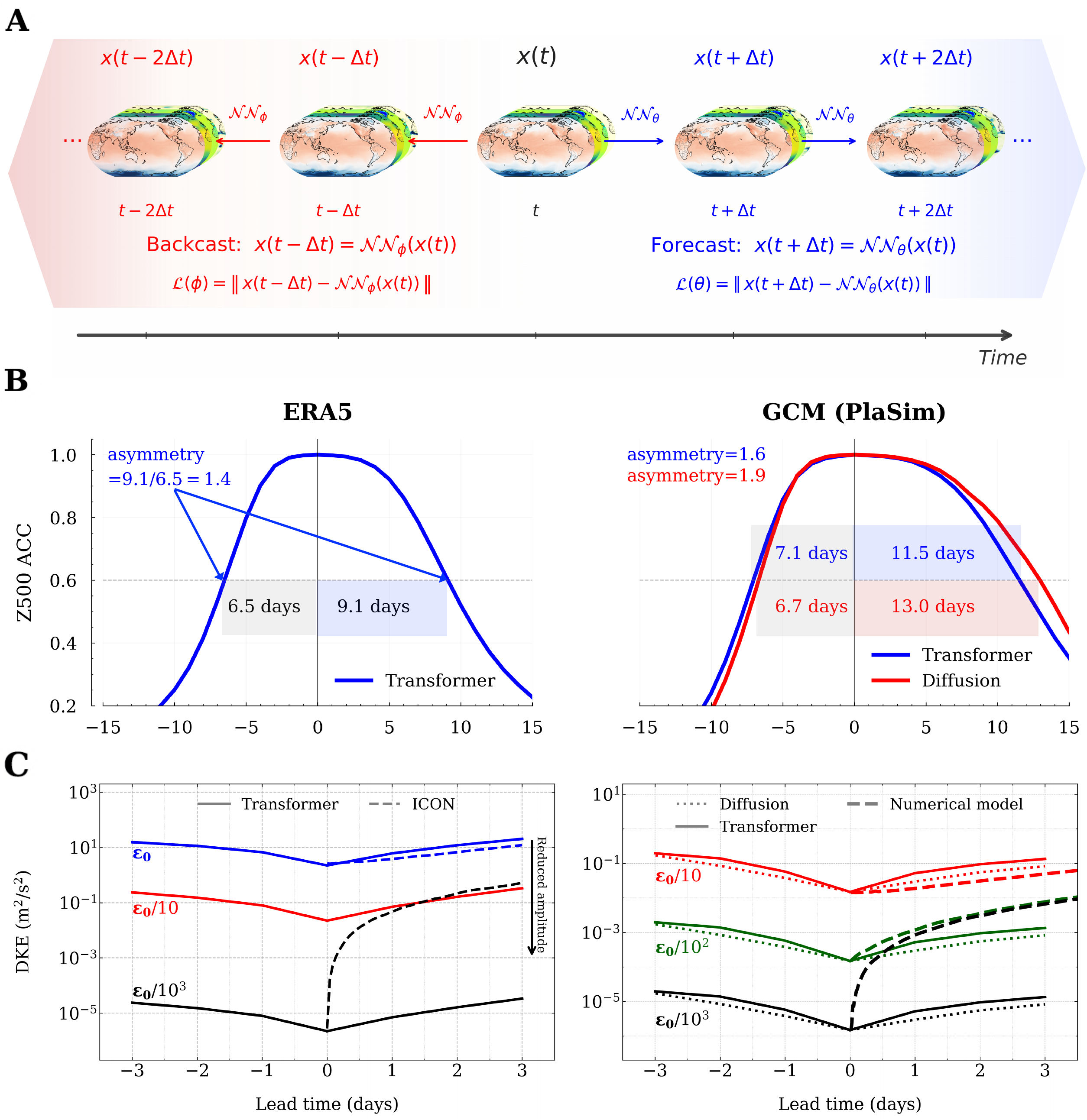} 
    \caption{
    \small{\textbf{Missing butterfly and forecast-backcast asymmetry in ERA5 and GCM AIWP models.}
    \textbf{(A)} Schematic of the autoregressive forecasting and backcasting models. The forecast network $\mathcal{NN}_{\theta}$ advances the atmospheric state  (Eq.~\eqref{eq:FC}); an independently trained backcast network $\mathcal{NN}_{\phi}$ steps it backward (Eq.~\eqref{eq:BC}).
    \textbf{(B)} Anomaly correlation coefficient (ACC) of Z500 versus lead time for ERA5 (left; Transformer (deterministic),  based on Pangu-Weather \cite{bi2023accurate}) and the PlaSim GCM (right; Transformer and conditional Diffusion (stochastic), similar to GenCast~\cite{price2025probabilistic}), averaged over all initial conditions from the test sets. $\Delta t=$~24~h and the horizontal resolution is $\sim 100$~km (ERA5) and $\sim 300$~km (PlaSim). Shading marks the ACC~$>0.6$ skill window; annotations give the crossing times and their forecast/backcast ratio, the asymmetry. See Figs.~\ref{fig:SI-SFNO}-\ref{fig:SI-RMSE-asymmetry} for other variables, metrics, $\Delta t$, and neural-operator and generative architectures. 
    \textbf{(C)} Difference kinetic energy (DKE) for perturbations of different amplitudes, decreasing from a reference value, $\epsilon_{0}$. We also show DKE from numerical integrations of physics-based models (ICON 20~km and PlaSim GCM). See Methods and Data for details about the AIWP models, physics-based models, perturbations, and metrics.}}
	\label{fig1:era5-plasim}
\end{figure}

Figure~\ref{fig1:era5-plasim}B shows the anomaly correlation coefficient (ACC) of 500-hPa geopotential (Z500) as a function of lead time. On ERA5, the backcasting Transformer (deterministic) remains skillful to 6.5~days into the past while the forecast is skillful for 9.1 days. It might be surprising that the past is harder to predict (asymmetry~$=9.1/6.5>1$). \textit{However, backcasting should be practically impossible for dissipative dynamical systems such as the atmosphere, because, simply put, it leads to an apparent violation of the second law of thermodynamics!} As further explained in the Supplementary Materials, time reversal turns dissipation (which is prevalent in the atmosphere across scales) into anti-dissipation (and reverses the sign of Lyapunov exponents), so unavoidable small errors grow explosively, fastest at the smallest scales. Accordingly, backcasting using numerical integrations fails catastrophically; see examples from the heat equation to Lorenz 63 to Lorenz 96 in the Supplementary Materials and Fig.~\ref{fig:rmse_hovmoller_1024}A. This is not the case for the AIWP model of ERA5. 

The asymmetry \(>1\) is not an artifact of ERA5, which, as an assimilation product continually updated by observations, is not a closed dynamical system: PlaSim, which is one (a GCM), exhibits even larger asymmetries, including in generative and neural-operator AIWP models (Figs.~\ref{fig1:era5-plasim}B and \ref{fig:SI-SFNO}-\ref{fig:SI-DiT}). The asymmetry is robust across variables,  pressure levels, and skill metrics in both ERA5 and PlaSim (Figs.~\ref{fig:SI-ACC-asymmetry}-\ref{fig:SI-RMSE-asymmetry}). Geographically, the asymmetry is overall larger in the tropics than in the extratropics (Table~\ref{tab:asymmetry}); likely a reflection of the stronger diabatic and dissipative processes (higher entropy production rate) there \cite{peixoto1991entropy,pauluis2002entropy}, so that the tropical atmosphere is, in this sense, further from reversibility (see the Supplementary Materials).

Two robust findings therefore need explaining: that backcasting works at all, and that the past is nevertheless consistently harder to predict than the future. But first, let us also examine the error growth and butterfly effect in these forecasting and backcasting AIWP models.

\subsection{Missing butterflies in AI forecasting and backcasting alike: ERA5 and PlaSim (GCM)}
In a multi-scale chaotic system like the atmosphere, ensemble spread from vanishingly small initial perturbations should grow rapidly, fed by upscale error transfer from the small scales, and saturate after a finite time. This is the ``real'' butterfly effect~\cite{lorenz1969predictability,palmer2014real,palmer2024real,sun2016intrinsic}, which sets the upper limit of atmospheric predictability. We quantify the spread with the difference kinetic energy (DKE), the area-weighted ensemble variance of the horizontal winds (Methods and Data). Figure~\ref{fig1:era5-plasim}C shows that numerical models, ICON (NWP) and the PlaSim GCM, reproduce this behavior; however, no AIWP model does, in forecasting or backcasting, for ERA5 or PlaSim.

The growth of DKE in the numerical integrations of physics-based forecast models behaves as expected: in both ICON and PlaSim GCM, the smallest-amplitude ensembles grow by orders of magnitude within a day or two, showing diminishing return as the initial-condition errors shrink (Fig.~\ref{fig1:era5-plasim}C). This is the butterfly effect \cite{palmer2014real,palmer2024real}. For the large-amplitude perturbations, ensemble forecasts from numerical integrations and AIWP models behave similarly, showing exponential growth in DKE over several days (unlike the butterfly effect, the growth rate here is independent of the perturbation amplitude). The forecasting results in ERA5 agree with those of Selz and Craig in their pioneering papers~\cite{selz2023can,selz2026can}. Here, we show the same behavior in the AIWP models of a GCM, but most strikingly, in backcasting AIWP models. The fundamental understanding of time-reversal physics (Supplementary Materials) and numerical experiments with Lorenz~96 (Fig.~\ref{fig:dke_1024}A) suggest even a \text{more} explosive growth of DKE in backcasting. However, we robustly see a near symmetry between the forecasting and backcasting DKE growth across the hierarchy.     

We are thus left with three puzzles about AIWP models: on one hand, forecast accuracy that long-standing arguments deemed unattainable, and on the other, skillful backcasting and a missing butterfly effect that disagree with the fundamental understanding of multi-scale chaotic dynamics, robustly confirmed by numerical integrations of physics-based models. To further probe these findings, we extend the hierarchy to a canonical testbed of multi-scale, chaotic dynamics, the Lorenz 96 system. 

\subsection{Training-data coarse-graining dramatically alters forecasts, backcasts, and butterflies: Lorenz 96}

\begin{figure}[t]
	\centering
	\includegraphics[width=\textwidth]{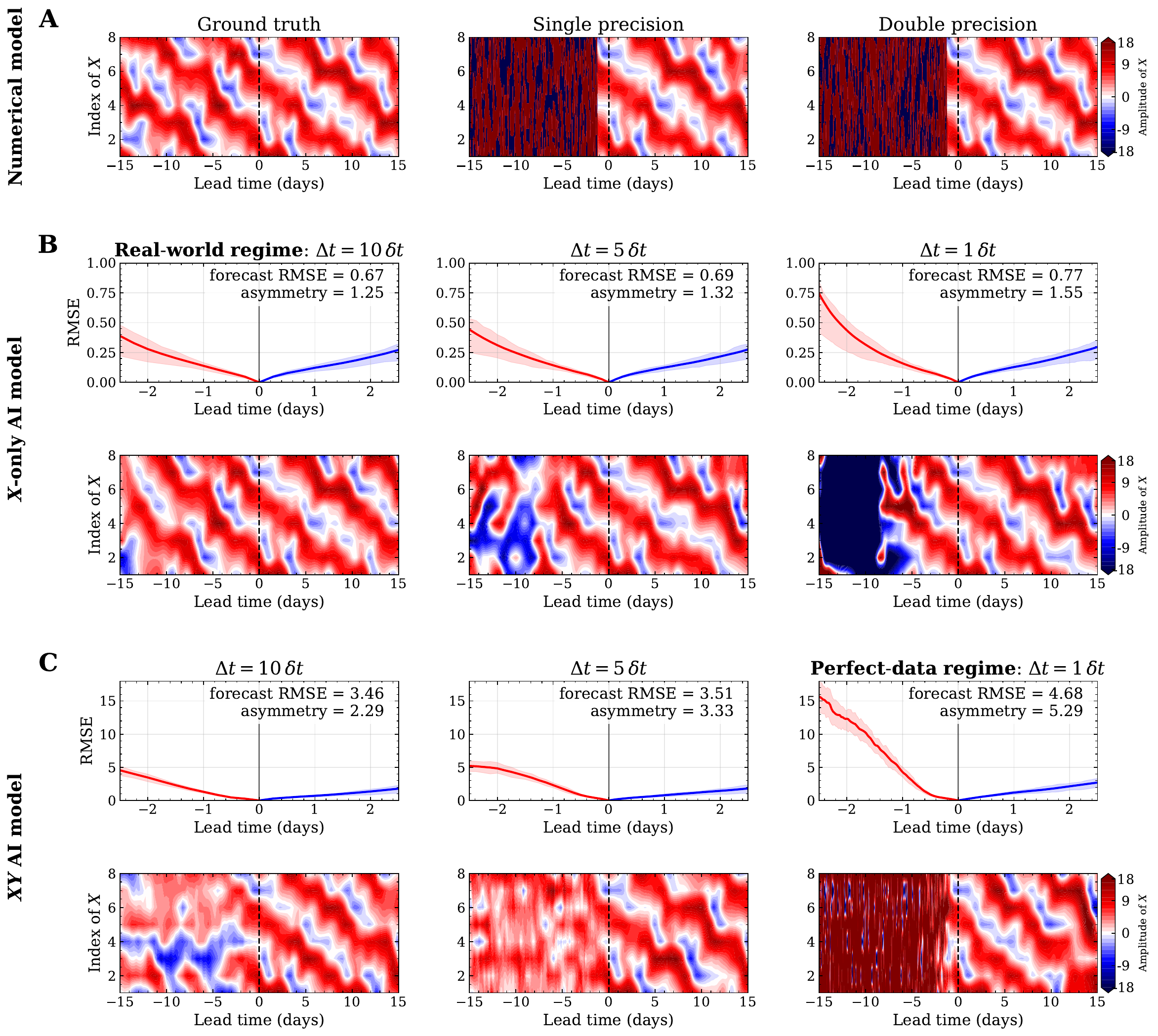} \caption{{\textbf{Effects of training data coarse-graining on forecasting and backcasting AI models of Lorenz 96.} All results (RMSE, asymmetry, and Hovm\"oller diagrams) are based on the large-scale variable, $X_i \, (i=1 \dots 8)$. (\textbf{A})~The numerical-model reference trajectory alongside forecasts and backcasts from numerical integrations of  Eqs.~\eqref{eq:L96_X}-\eqref{eq:L96_Y} initialized at lead time $0$; saturated dark red/blue regions mark numerical blow-up in backcasts. Single floating-point precision calculations are shown to match the precision used in AI models' inference; it does not make any noticeable difference. (\textbf{B}) and (\textbf{C}) show RMSE curves (averaged over 100 initial conditions from the test set) and Hovm\"oller diagrams (from the same initial condition as in (A)) for 12 independent AI models trained on only the large-scale $X_i$ and on both scales ($X_i$ and $Y_{i,j} \, (j=1 \dots 32$)) with time interval $\Delta t = 10\delta t$, $5\delta t$, and $1\delta t$ ($\delta t$ is the time step of the numerical solver). Shading indicates the RMSE interquartile range. Annotation boxes report forecast RMSE at 4.5~days and the asymmetry (backcast/forecast RMSE ratio) at 1.5~days. Note the difference between the RMSE ranges in rows (B) and (C), as well as between the lead-time ranges, chosen for clearer illustration. See Methods and Data for more details of the system and AI models.}}
    \label{fig:rmse_hovmoller_1024}
\end{figure}

\begin{figure}[t]
	\centering
	\includegraphics[width=\textwidth]{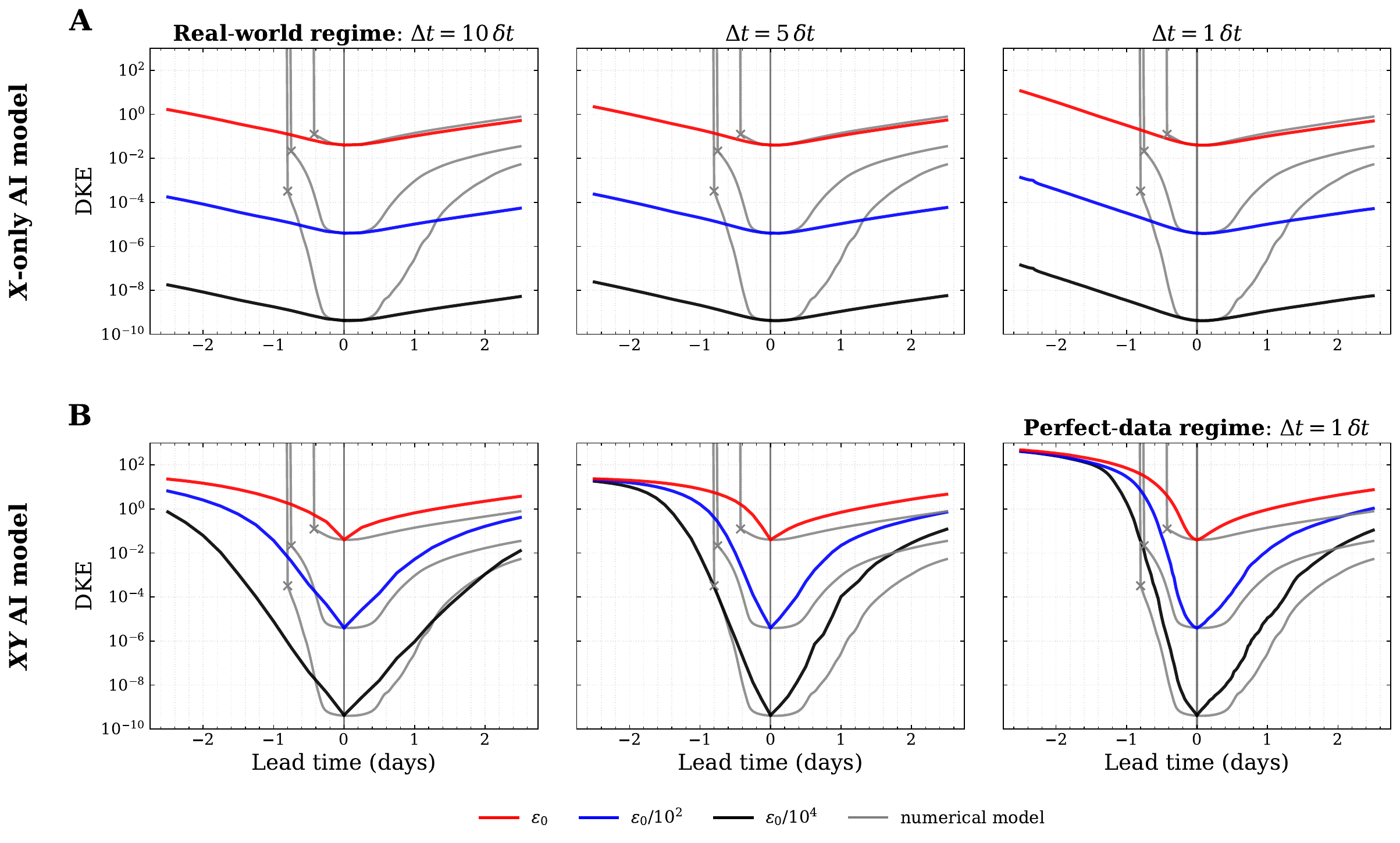} 
	\caption{{\textbf{Effects of training-data coarse-graining on ensemble growth and the butterfly effect in AI models of Lorenz 96.} Curves show DKE of the large-scale variable, $X_i \, (i=1 \dots 8)$ for 12 independent, separate AI models that have an identical architecture (including the same MSE loss) trained on (\textbf{A}) only the large-scale $X_i$ and (\textbf{B}) on both scales  ($X_i$ and $Y_{i,j} \, (j=1 \dots 32$)) with time interval $\Delta t = 10\delta t$, $5\delta t$, and $1\delta t$ ($\delta t$ is the time step of the numerical solver). Gray curves show forecasts and backcasts from numerical integrations of the two-scale Lorenz 96 equations (\ref{eq:L96_X})--(\ref{eq:L96_Y}); crosses mark numerical blow-up in backcasts. Curves are averaged over the same 100 initial conditions from the test set. Perturbations are of different amplitudes, decreasing from a reference value, $\epsilon_{0}$. See Methods and Data for details about the perturbations and metrics.}}
	\label{fig:dke_1024}
\end{figure}

The two-scale Lorenz 96 system (Eqs.~\eqref{eq:L96_X}-\eqref{eq:L96_Y}) couples 8 slow, large-amplitude variables $X_i$ (representing large-scale circulation) to 256 fast, small-amplitude variables $Y_{i,j}$ whose characteristic time scale is an order of magnitude shorter (representing small-scale, often subgrid-scale processes); see Methods and Data. Its numerical integration displays the time irreversibility that should make any backcasting impractical: Fig.~\ref{fig:rmse_hovmoller_1024}A shows that integrated backward, the trajectory blows up rapidly due to anti-dissipation. As demonstrated in Fig.~\ref{fig:dke_1024}, the numerical ensemble forecasts of this system exhibit the butterfly effect, and the overall DKE growth curves show a behavior similar to those from numerical integration of physics-based models like ICON and PlaSim GCM as the perturbation amplitudes are varied (compare to Fig.~\ref{fig1:era5-plasim}C). 

The AI forecasting and backcasting models can exhibit very different behaviors depending on the degree of \textit{coarse-graining} of the \textit{training data}: How many spatio-temporal scales the training data includes. We first train AI models with data closest to the ``real-world'' regime: only using the large scales and after removing the fastest variabilities ($X$-only, $\Delta t=10 \delta t$ where $\delta t$ is the time step of the numerical solver/ground truth). As shown in the top-left corners of Figs.~\ref{fig:rmse_hovmoller_1024}B-C and \ref{fig:dke_1024}, the forecasts and backcasts of these AI models have remarkably similar characteristics as those of AIWP models of ERA5 and PlaSim (Figs.~\ref{fig1:era5-plasim}B-C and \ref{fig:SI-ACC-asymmetry}-\ref{fig:SI-RMSE-asymmetry}): Accurate forecasts, skillful backcasts with asymmetry $>1$, amplitude-independent exponential growth in DKE, and missing butterflies. Hovm\"oller diagrams and DKE curves of backcasts demonstrate behavior that is markedly different from that of numerical integrations. 

Note that the missing butterfly effect in the real-world-regime AI model of two-scale Lorenz 96 and in the AIWP models of ERA5 and PlaSim (Figs.~\ref{fig1:era5-plasim}C and \ref{fig:dke_1024}) does not mean that these models miss chaos. Their predictions are sensitive to initial-condition perturbations across perturbation amplitudes. In fact, their amplitude-independent exponential DKE growth closely resembles that of numerical integrations of Lorenz~63 (Fig.~\ref{fig:SI-L63-DKE}), the hallmark example of deterministic chaos, but itself misses the real butterfly effect because it lacks multi-scale dynamics (see Palmer~\cite{palmer2014real, palmer2024real} for an illustrative discussion of Lorenz~63 versus Lorenz~96).

Next, we develop additional (independent) forecasting and backcasting AI models trained on data with more spatial and/or temporal scales: using $X$ or $(X,Y)$ and $\Delta t=1$, 5, or 10\,$\delta t$ (see Methods and Data for details of extensive model design exploration). At the bottom-right corner of Figs.~\ref{fig:rmse_hovmoller_1024}B-C and \ref{fig:dke_1024}, we show the results of the AI models trained in the ``perfect-data'' regime: these models have seen all the scales the numerical solution/ground truth contains. Compared with the models trained in the ``real-world'' regime, the AI models trained in the perfect-data regime have a worse forecast skill (by a factor of $\sim 7$) and larger asymmetry (by a factor of $\sim 4$) but also blown-up backcasting and rapid DKE growth that closely mimics the butterfly effect. The last two in particular show that the perfect-data-regime AI models are consistent with the fundamental understanding of multi-scale chaotic systems and their numerical integrations. 

We observe these trends in forecast skill, asymmetry, DKE growth, and Hovm\"oller diagrams consistently as we move from the ``real-world'' regime corner to the ``perfect-data'' regime corner. Therefore, we propose the following three hypotheses supported by these Lorenz 96 experiments: \textit{coarse-graining} used in producing the \textit{training data} is the common cause of H1) Unexpected AI forecast skill (for large scales), H2) Missing butterfly effect, and H3) Skillful backcasting. We broadly define coarse-graining, which is inevitable in practice, as filtering of small/fast scales and/or ignoring some state variables. 

The emergence of H1 from coarse-graining can be further seen in the analysis of leading finite-time Lyapunov exponents of the large and small scales, $\lambda^X$ and $\lambda^Y$ (Table~\ref{tab:si_lyapunov}). In the ground truth, $\lambda^X$ and $\lambda^Y$ are the same (within the uncertainty range), reflecting the fact that the error growth in the large scales is determined by the small scales. The perfect-data-regime AI model reproduces these Lyapunov exponents fairly well. However, the real-world-regime AI model, which lacks small scales, has $\lambda^X$ around 8 times smaller than the ground truth/perfect-data-regime AI model, leading to its enhanced forecast skills. We note that here, H1 refers to the skill for large scales, which is relevant for practical forecasting and the basis of common verification metrics (see Discussion).   

As for H2, Fig.~\ref{fig:dke_1024} shows the dramatic impact of training-data coarse-graining on the DKE growth of the 12 AI forecasting and backcasting models, which share an identical architecture and MSE loss. In particular, when trained on complete or nearly complete data, these MSE-trained models exhibit a butterfly-like effect: rapid, amplitude-dependent DKE growth comparable to, or even faster than, that of the numerical integrations (Fig.~\ref{fig:dke_1024}B). This argues against the suggestion that optimizing toward a conditional mean is the primary cause of the missing butterfly effect (see the Discussion). These results raise the question of whether a butterfly-like effect can likewise emerge in a state-of-the-art AIWP model when the coarse-graining of its training data is reduced, which we address next. 

\subsection{As coarse-graining is reduced, a butterfly-like effect emerges and forecast skill degrades: Pangu-Weather}

\begin{figure}[t]
	\centering
	\includegraphics[width=1.0\textwidth]{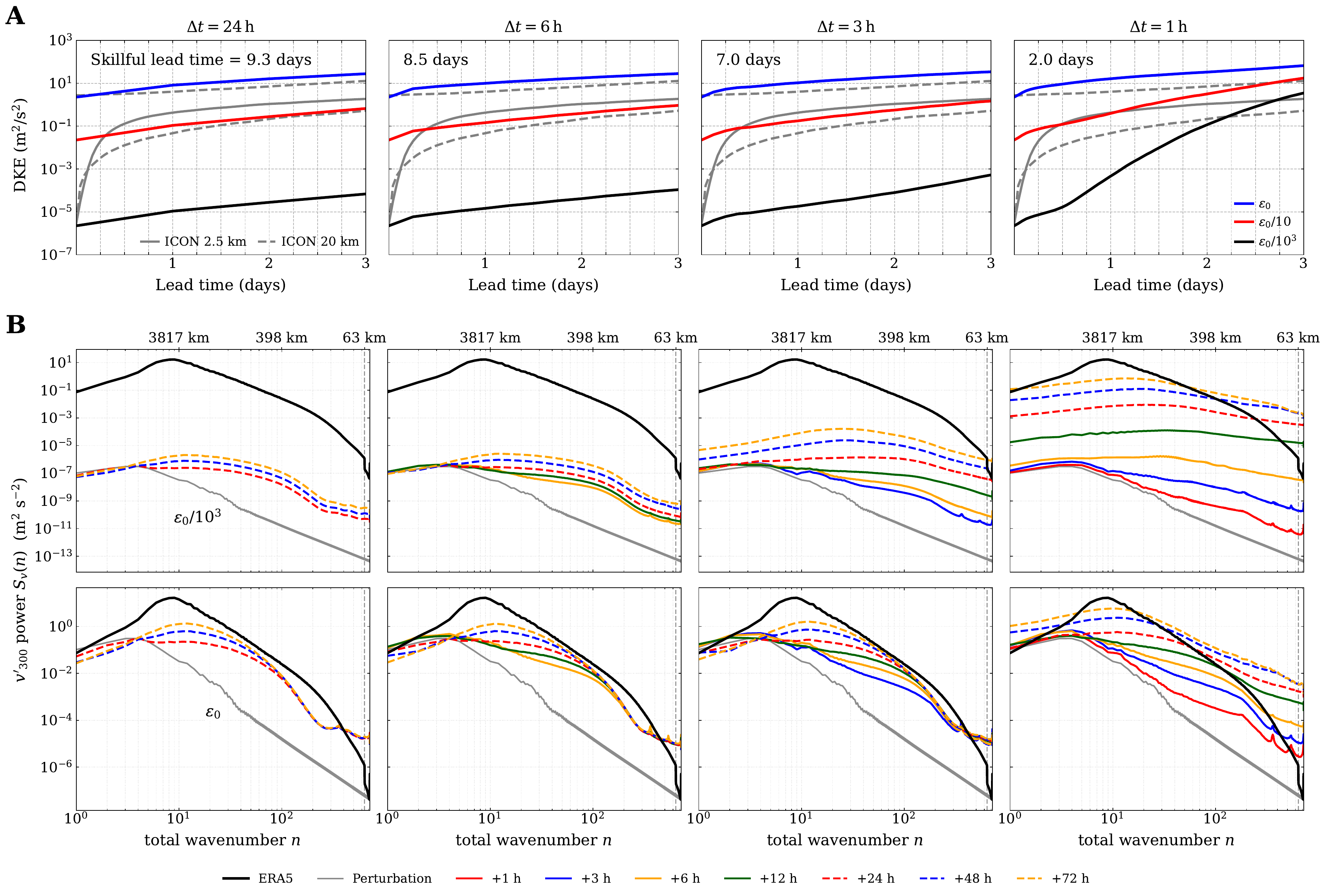}
	\caption{{\small \textbf{Effects of temporal coarse-graining of training data on ensemble growth and the butterfly effect in the Pangu-Weather forecasts.}
	Columns show, from left to right, forecasts from the independent, separate official Pangu-Weather models \cite{bi2023accurate} with time step $\Delta t = 24$, 6, 3, and 1~h. 
	(\textbf{A}) DKE versus lead time, as in Fig.~\ref{fig1:era5-plasim}C, for three amplitudes of initial perturbations. DKE is averaged over $60^\mathrm{o}$S-$60^\mathrm{o}$N. See Fig.~\ref{fig:SI-sensitivity-Fig4-tf32-global} for results with global averaging and lower GPU precision. Gray curves show the DKE from ICON forecasts (from \cite{selz2023can}). The number in each panel is the skillful lead time, i.e., when ACC drops to 0.6, as in Fig.~\ref{fig1:era5-plasim}B.
	(\textbf{B}) Power spectra $S_{v}(n)$ of the 300-hPa meridional wind perturbation $v'_{300}$ (each member minus ensemble mean) as a function of total wavenumber $n$, for the initial-condition perturbations with the smallest ($\epsilon_{0}/10^{3}$, upper row) and largest ($\epsilon_{0}$, lower row) amplitudes. These curves are averaged over the same ensemble members and initial conditions as the ERA5 Transformer in Fig.~\ref{fig1:era5-plasim}C. Colors denote (nonuniform) lead times from $+1$ to $+72$~h; the gray curve is the imposed initial Perlin noise perturbation at $t = 0$, and the black curve is the spectrum of the full ERA5 field for reference. The top axis denotes the corresponding wavelength. Figure~\ref{fig:SI-error-spectra} shows the spectra of the forecast errors as a function of lead time. See Methods and Data for more information.}}
	\label{fig:downsampling}
\end{figure}

We use the official Pangu-Weather models \cite{bi2023accurate}: four independent deterministic neural networks with time steps $\Delta t = 24$, $6$, $3$, and $1$~h, all trained with an MAE loss on $0.25^\circ$ ERA5, the native spatial resolution of the dataset. We only perform inference (forecasts) with these models; we have not trained, fine-tuned, or otherwise modified them. Figure~\ref{fig:downsampling}A shows that as $\Delta t$ decreases, the small-amplitude ensembles
transition from slow, \textit{amplitude-independent} exponential DKE growth for the $24$-h Pangu-Weather, consistent
with \cite{selz2023can,selz2026can} and our $1^\circ$ Transformer (Fig.~\ref{fig1:era5-plasim}C), to rapid, \textit{amplitude-dependent} growth for the $1$-h Pangu-Weather that approaches the butterfly-like behavior of the ICON integrations. Meanwhile, the skillful lead time drops from 9.3 to 2.0 days. Consistent with the Lorenz~96 experiments (Figs.~\ref{fig:rmse_hovmoller_1024}-\ref{fig:dke_1024}) and hypotheses H1 and H2, reducing (here, temporal) coarse-graining degrades forecast skill and strengthens butterfly-like growth.

This behavior has not been reported before because the $\Delta t=1$~h and $3$~h Pangu-Weather models had not been run continuously. Selz and Craig~\cite{selz2023can} used all four networks, but through hierarchical temporal aggregation, in which the $\Delta t=24$~h model takes the longest steps and the shorter-$\Delta t$ models only fill the intermediate hours; the drops in DKE at network switches that they reported, largest when the 24-h model takes over, are consistent with Fig.~\ref{fig:downsampling}A. Their follow-up study \cite{selz2026can} ran only the 6- and 24-h versions.

The spectra of the ensemble perturbations (Fig.~\ref{fig:downsampling}B) further characterize the emerging DKE growth. For $\Delta t = 24$ and $6$~h, the small-amplitude spectra jump within the first step and then grow slowly, with their shape roughly preserved, remaining orders of magnitude below the ERA5 spectrum
even at 3 days. For $\Delta t = 1$~h, perturbation energy instead grows fastest at the small scales, saturates there first, and progressively fills in the larger scales toward the ERA5 spectrum: the upscale error growth expected in multi-scale chaotic systems \cite{lorenz1969predictability,palmer2014real}; see Fig.~\ref{fig:SI-error-spectra}. However, at the smallest scales, the late-time spread overshoots the ERA5 spectrum, beyond the factor of 2 allowed by physical saturation; the 1-h model generates variance at scales that ERA5 itself barely contains, a warning sign discussed in Methods and Data.

We emphasize that we are not claiming that this is the (real) butterfly effect. Even at its native resolution, ERA5 (and especially its subset of variables used for training) is highly coarse-grained: the reanalysis is produced by a $\sim$31-km NWP model with parameterized subgrid processes, it misses state variables, and $\Delta t = 1$~h remains far longer than the time step of the NWP model's numerical solver. The Lorenz~96 hierarchy shows what to expect in this situation: with small spatial scales fully missing ($X$-only), AI models overall remain in the real-world-regime behavior even when trained at the solver's own time step (Fig.~\ref{fig:dke_1024}, top-right corner, where $\Delta t = 1\delta t$). Even in the perfect-data regime, we call the behavior in Fig.~\ref{fig:dke_1024} \textit{butterfly-like}: the DKE grows faster than the numerical model's near $t=0$, whereas the real-world-regime AI models match the early numerical growth rates better but never accelerate as time progresses forward or backward. Returning to the Pangu-Weather model with $\Delta t=1$~h, unlike what numerical experiments show~\cite{zhang2003effects, selz2015upscale, selz2023can}, the largest early DKE growth is not necessarily collocated with precipitation (Fig.~\ref{fig:SI-DKE-map+precip}). However, we offer a word of caution and argue that assessing the butterfly effect, and more broadly, the physicality of error growth, in AI models requires deeper investigation and a revisiting of what these terms mean, as the current understanding heavily relies on numerical experiments, with their own known shortcomings. Disagreement with a numerical model does not, by itself, make an AIWP model wrong: these data-driven models might be learning, and forecasting, weather differently.

One must also be cautious about hardware artifacts \cite{selz2026can} and nonphysical forecasts while examining butterflies in AI models; see Methods and Data for how these are controlled for and assessed. What is robust here is the trade-off: within a single family of state-of-the-art AIWP models, reducing temporal coarse-graining results in butterfly-like growth and degrades forecast skill, further supporting H1 and H2.

\section{Discussion}
Across the hierarchy of ERA5, PlaSim GCM, and Lorenz 96, we find that the \textit{coarse-graining} of the \textit{training data} is the common cause of the following (seemingly unrelated) key features of AI prediction models:
\begin{enumerate}
    \item [H1.] The surprising forecast accuracy,
    \item [H2.] The missing butterfly effect,
    \item [H3.] The skillful backcasting.
\end{enumerate}
In short, the more completely the training data represent the true dynamical system spatially and temporally, the more the AI model behaves like the system, but the worse it forecasts the large scales, which are often the desired target of prediction and most verification metrics.

Before discussing H1-H3 and their implications, we note that these results offer a step toward answering a central question about AIWP models: Do they learn atmospheric dynamics, or merely learn the evolution of weather the way one learns a video? A video plays backward as easily as forward. Skillful backcasting (H3), taken alone, might appear to support this video interpretation, but only in the strongly coarse-grained regime characteristic of the AIWP models' common training datasets. As the training data become more complete, the same architecture, trained with the same loss on pairs one time step apart, becomes markedly more physics-like: backcasting skill disappears, the forecast--backcast asymmetry grows, and a butterfly-like effect emerges. We are not suggesting that AIWP models trained on strongly coarse-grained data (e.g., ERA5 with $\Delta t=24$~h) learn weather much like a video: dynamical tests and other analyses have demonstrated that they learn aspects of atmospheric dynamics \cite{sun2025can, sun2025predicting, van2025reanalysis, wu2025applying, hakim2024dynamical,  craig2026physics}. Our findings, however, show that with more complete training data, these models behave more like the physical system, including its arrow of time, although they become less accurate for practical forecasting. A major question then is whether there is a downside to having AIWP models that are more accurate but less physics-like; we return to this later. 

Regarding H1, coarse-graining removes the fast, small scales through which errors grow upscale to the large scales \cite{lorenz1969predictability, sun2016intrinsic, zhang2003effects, selz2015upscale}. Therefore, the \textit{effective system} that the AI models learn has much slower large-scale error growth than the true system (quantified for Lorenz 96 via Lyapunov exponents; Table~\ref{tab:si_lyapunov}). That lower spatio-temporal resolution improves forecast accuracy might seem counterintuitive, but this intuition is built on numerical simulations, in which higher resolution improves accuracy. This is because of lower discretization errors and, more importantly, reduced reliance on ``explicit'' subgrid-scale parameterization schemes, the leading source of structural/epistemic uncertainty in these models. AI models, in contrast, learn subgrid-scale parameterizations ``implicitly'': they can account for the averaged effects of the fast, small scales on the large scales' evolution without explicitly representing the former, which can otherwise drive rapid error growth. Consistent with this implicit parameterization, AI climate emulators, trained on coarse-grained outputs of high-resolution physics-based simulations can closely reproduce key climate statistics of the original simulations \cite{perkins2025hiro}. Moreover, even in NWP models, increasing resolution yields diminishing returns in large-scale forecast accuracy \cite{buizza2010horizontal}, in part because the newly resolved fast, small scales drive rapid error growth \cite{zhang2019predictability, selz2015upscale, judt2018insights}. Taken together, we argue that this \textit{implicit parameterization} in AIWP trained with coarse-grained data significantly enhances their forecast skill and is a major reason for their surprising accuracy.   

We should clarify that the surprising accuracy is not necessarily about comparison with the NWP models. Such a comparison would have to account for the quality of the physics-based models' subgrid-scale parameterization schemes, a complex question. Rather, H1 concerns the AIWP models' skill, which is unexpected given earlier studies arguing that data-driven weather forecasting requires an enormous amount of training data \cite{lorenz1969atmospheric}, with $10^{30}$ years as one estimate~\cite{van1994searching}. Our findings suggest that such estimates need to be revisited in light of the effects of AIWP models' training-data coarse-graining on error growth and the butterfly effect, key contributors to those high data requirements. Other factors likely contribute to the gap between these estimates and practice, including
the underlying assumptions, e.g., about the attractor dimension \cite{nicolis1998atmospheric, platzer2021using}, and, perhaps more importantly, the learning algorithm. Analog forecasting in those studies is essentially nearest-neighbor regression, whose scaling suffers from the curse of dimensionality \cite{stone1982optimal,gyorfi2002distribution}. Current AIWP models instead use deep neural networks (over-parameterized nonlinear regression) to learn a smooth  $\Delta t$ flow map \cite{barron1993universal,poggio2017and, belkin2019reconciling}. Furthermore, AIWP models have been shown to “translocate”, i.e., learn from dynamically similar events in other regions, which significantly enrich their training set \cite{sun2025predicting,sun2025can}. Developing theoretical scaling laws relating training-data size to AIWP model accuracy, beyond emerging empirical estimates \cite{yu2026scaling,subramanian2026neural}, remains an important open research direction and can build on our findings. 

As for the missing butterfly effect (H2), it is a symptom of coarse-grained training data, not of the architecture, the loss function, or deterministic-versus-generative design. The Lorenz 96 experiments isolate the cause cleanly (Fig.~\ref{fig:dke_1024}): all 12 networks share one architecture and one MSE loss, yet the butterfly effect appears or disappears with the content of the training data alone. The Pangu-Weather experiments with varying $\Delta t$ (Fig.~\ref{fig:downsampling}) show the same dependence in a state-of-the-art AIWP model. Spatially, however, ERA5 (and PlaSim) are already highly coarse-grained at their native resolutions, leaving no room to informatively further vary the spatial scales in either direction; future explorations should focus on high-resolution global NWP models and GCMs. These findings argue against optimizing toward a conditional mean or blurring, which is due in part to spectral bias \cite{chattopadhyay2023challenges}, as the primary cause: MSE-trained networks reproduce the butterfly effect in Lorenz~96, while the Diffusion AIWP models of PlaSim studied here and GenCast and FourCastNet2 examined in \cite{selz2026can} miss butterflies despite their little spectral bias (Figs.~\ref{fig1:era5-plasim}C,~\ref{fig:SI-DiT}, and ~\ref{fig:SI-spectra-Bias-for-different-dt}). 

Note that after showing the lack of butterfly effect across AIWP models of different architectures and loss functions, \cite{selz2026can} suggested a training-data origin, the remaining common factor (although all these AIWP models also share the same learning principle: supervised prediction of the short-$\Delta t$ evolution). However, the problem has been attributed to the specific nature of the data-assimilated analysis products used for training and fine-tuning~\cite{selz2026can,palmer2024real}. The hierarchy here, whose PlaSim and Lorenz 96 training sets involve no data assimilation, tests the training-data origin directly and identifies the key property: its spatial and temporal coarse-graining. Our results also suggest that restoring butterflies can come at the cost of forecast accuracy; still, if one views their absence as a bug rather than a feature of accurate AIWP models, the question becomes how best to restore them (see below).

Skillful backcasting (H3) is made possible by the same coarse-graining. The learned effective system is smoother, with smaller leading Lyapunov exponents and less of the fast, small scales that dominate the anti-dissipative directions responsible for explosive backward integration (see Supplementary Materials). The apparent violation of the second law of thermodynamics is thereby resolved: the learned effective system is less irreversible than the true one, but still irreversible. This residual irreversibility is manifested in the asymmetry remaining above one, which is largest in the tropics, consistent with the higher entropy production there~\cite{peixoto1991entropy,pauluis2002entropy}, and in preliminary explorations in which long-term AI backcasting of Lorenz~96, PlaSim, and ERA5 leads to unstable or unphysical states, a sign of irreversibility, though much slower than the true system. 

This work introduces backcasting as a new, useful lens for investigating the fundamentals of predictability and better understanding AIWP models. Backcasting can also provide practical tools of its own. Speculatively, a single AI model trained both forward and backward might internalize more of the underlying dynamics, perhaps even rudiments of causality. That said, our preliminary bidirectional forecasting-backcasting model (Methods and Data) shows no improvement so far, but the design space is essentially unexplored. Backcast models might also help sample the rarest weather extremes and find their dynamical precursors, tasks traditionally addressed with limited linear tools such as adjoints. AI-based approaches have recently emerged, including AI-boosted rare-event sampling \cite{lancelin2026ai}, iterative use of AIWP models' adjoints \cite{vonich2024predictability}, and sampling emulators like cBottle  \cite{brenowitz2025climate, lin2026extreme}. Backcasting could augment these as a stable, fully nonlinear, autoregressive counterpart of the adjoints.

Growing efforts focus on identifying the physics that AI models miss \cite{sun2025can,zhang2026physics, chattopadhyay2023challenges, bonavita2024some, selz2023can,selz2026can, kim2026spectral} and on making them more physics-like, e.g., through inductive biases such as constraints in architectures and loss functions \cite{lai2024machine, li2024physics, verma2024climode, sha2025improving}. Our results prompt a question: do we actually want AIWP models to fully behave like physics-based models even if restoring the physics costs performance, e.g., accuracy of practical forecasting (Figs.~\ref{fig:rmse_hovmoller_1024} and \ref{fig:downsampling})? This trade-off may prove solvable, and there is precedent for excluding physics for the benefit of practical skill: sound waves, computationally costly to resolve yet largely irrelevant to weather, have been filtered out of the equations solved by many NWP models and GCMs \cite{vallis2006atmospheric}. However, the distinction is that while sound waves carry little of what matters for weather and climate prediction, the fast, small scales filtered by coarse-graining carry much of it. Thus, the deeper question is what these AI models are missing because coarse-graining has removed physics from their training data. Is this absence part of why AIWP models struggle with gray swans \cite{sun2025can,zhang2026physics}? What does it imply for probabilistic forecasts, especially at subseasonal-to-seasonal lead times, which rely on perturbation growth and must be well calibrated \cite{aitken2026designing,asch2026rigorous}? The question might be even more pressing for long-term climate emulators: AI models of the emerging global km-scale, physics-based simulations \cite{stevens2019dyamond} are typically trained on output first heavily coarse-grained, e.g., to $\sim$100~km \cite{perkins2025hiro}, to $\Delta t$ of hours to a day, and even to daily-averaged state variables \cite{henn2026aimip}, which can discard significant physics. Consistent with this concern, idealized data-driven models with partial state vectors (missing variables) fail to reproduce forced responses \cite{falasca2025probing}. Such failures offer a caution for emulators built for sampling internal variability and projecting climate change \cite{lai2024machine}. Answering these questions requires a deeper understanding of what AI weather and climate models learn, do not learn, and why.

To conclude, the development of AI weather (and climate) models has concentrated on architectures and objectives: deeper and larger networks, customized loss functions, and embedded physical constraints. Those directions address the optimization and expressiveness of AI models. Our results point instead to the ingredient that no architecture or loss function can substitute for: the information content of the training data. Two remedies might come to mind immediately: hybrid AI-physics modeling, and foundation modeling, in which heterogeneous datasets are combined through large-scale pretraining. However, state-of-the-art examples of each, NeuralGCM \cite{kochkov2024neural} and Aurora \cite{bodnar2025foundation}, miss the butterfly effect \cite{selz2026can}; restoring at least this feature appears to require other approaches. Coarse-graining does entail an irreversible loss of information, yet some of what is lost can, in principle, be represented rather than resolved, as memory and stochasticity: the Mori--Zwanzig formalism \cite{mori1965transport, zwanzig1960ensemble} and its data-driven descendants \cite{wouters2013multi,kondrashov2015data,falasca2026physics} (see Supplementary Materials) might provide a principled starting point. For predicting the large scales of weather days ahead, coarse-graining is a feature, the source of AIWP skill and of learnable backcasting; for representing the physics of the atmosphere, its error growth, irreversibility, and predictability limits, it is a bug, likely most consequentially for long-term emulation. The content of the training data is therefore not a preprocessing detail but the central design decision of an AI weather or climate model.

\clearpage

\begin{ack}
We thank Kerry Emanuel and Fabrizio Falasca for insightful discussions about atmospheric predictability and time reversal, and Boris Bonev and Daniel Boscu for helpful comments on GPU precision. Computational resources were provided by NSF ACCESS (allocation ATM170020), NCAR’s CISL (allocations URIC0009 and UCHI0018), and the University of Chicago's Research Computing Center and Data Science Institute (DSI). Claude Fable 5 has been used for editing and proofreading the text, and for generating the visualization code for Movie~S1.

\paragraph*{Funding:}
This work was supported by NSF grant AGS-2531264, Schmidt Sciences LLC (through the InMOS project), and the University of Chicago’s Data Science Institute (DSI) and the Institute for Climate and Sustainable Growth (through the AI for Climate Initiative) to PH. AW and JF are grateful to DSI for an Eric and Wendy Schmidt AI in Science and an AI for Climate postdoctoral fellowship, respectively. JQW was supported by a Pritzker AI+Science Visiting Scholarship from DSI. 

\paragraph*{Author contributions:}
PH conceived the idea and wrote the paper. PH and YQS supervised research. PH, YQS, and JQW designed the study. WL and YQS (ERA5 and PlaSim) and JW (Lorenz systems) trained AI models, analyzed data, generated the figures, and wrote the Methods and Data section. AW developed AI models and generated the PlaSim data. All authors interpreted the results and edited the paper.

\paragraph*{Competing interests:}

There are no competing interests to declare.
\end{ack}

\paragraph*{Data and code availability:}

We downloaded the ERA5 dataset from Copernicus \url{https://cds.climate.copernicus.eu/} and regridded it with CDO \url{https://code.mpimet.mpg.de/projects/cdo/}. The ICON data are provided by~\cite{selz2023can}. The PlaSim GCM is available at \url{https://github.com/HartmutBorth/PLASIM}.  The official Pangu-Weather models are obtained from \url{https://github.com/198808xc/Pangu-Weather}. Movie S1 is available at \url{https://doi.org/10.5281/zenodo.22062019}. We will make all trained AI models and their inference data public upon publication.

\newpage

\bibliographystyle{unsrtnat}
\bibliography{FB_ref}

\newpage

\appendix

\renewcommand{\thefigure}{S\arabic{figure}}
\renewcommand{\thetable}{S\arabic{table}}
\renewcommand{\theequation}{S\arabic{equation}}
\renewcommand{\thepage}{S\arabic{page}}
\setcounter{figure}{0}
\setcounter{table}{0}
\setcounter{equation}{0}
\setcounter{page}{1}

\section{Methods and Data}

Throughout the Methods and Data, $\delta t$ denotes the numerical integration time step used to generate training and testing data (in PlaSim and Lorenz 96), $\Delta t$ denotes the input--output prediction interval used to train and infer with an AI model, and $\tau$ denotes the forecast or backcast lead time. With the exception of Fig.~\ref{fig:SI-sensitivity-Fig4-tf32-global}C, all results and analyses presented in this paper are from AI models trained and inferred with single precision (FP32) on the NCAR CISL Derecho cluster using A100 GPUs (more details below). \\

\noindent We provide details of the following in this section:
\begin{itemize}
    \item The ERA5 dataset and the AI models trained on it,
    \item The ICON NWP model and data,
    \item The PlaSim GCM, the dataset generated from it, and the AI models trained on this data,
    \item The two-scale Lorenz 96 system, the dataset generated from it, and the AI models trained on this data, 
    \item The evaluation metrics. \\
\end{itemize}

\noindent The Supplementary Materials includes 
\begin{itemize}
    \item Discussion on time reversal in multi-scale, dissipative dynamical systems,
    \item Movie S1 visualizing the 3D phase space of numerical forecasts and backcasts in Lorenz~63, 
    \item More details on the Transformer-, SFNO-, and Diffusion-based AI model architectures.
\end{itemize}

\subsection{Observation-based data: ERA5}

\noindent\textit{Data}\\
ERA5~\cite{hersbach2020era5} is the fifth-generation global atmospheric reanalysis of the European Centre for Medium-Range Weather Forecasts (ECMWF) and serves as the observationally constrained atmospheric dataset. We use five upper-air variables on 17 pressure levels and a selection of surface variables as our prognostic state $\mathbf{x}$ (see Table~\ref{tab:variables} for details). Each AIWP model evolves the prognostic variables in time conditioned on prescribed time-varying forcings $\mathbf{b}(t)$ and static boundary fields $\mathbf{c}$, which are supplied as inputs but are not evolved. 

As described below, we have also trained independent, separate forecasting and backcasting AIWP models with a transformer architecture that is based on Pangu-Weather. These deterministic models (Transformers, hereafter), used in Fig.~\ref{fig1:era5-plasim}, are trained on data remapped conservatively to a uniform $1^{\circ}\times1^{\circ}$ latitude--longitude grid. We have trained independent, separate models with $\Delta t=6$ and $24$~h. We train on 6-hourly data from 1979--2018 (40 years) and validate and select checkpoints on 2019. 

The official Pangu-Weather model, which we obtained from~\cite{bi2023accurate}, is trained on data with $0.25^{\circ} \times 0.25^{\circ}$ latitude--longitude grid; there are independent, separate versions trained with $\Delta t=1, 3, 6$ and $24$~h. We have only performed inference for ``forecasting'' with the Pangu-Weather model (Fig.~\ref{fig:downsampling}).

These Transformer AIWP models and the Pangu-Weather model are tested using the same initial conditions taken once a month in 2020 and 2021.\\

\noindent\textit{Forecasting and backcasting AIWP models: Transformer}\\
The forecast and backcast AI models are formulated based on Eqs.~\eqref{eq:FC}--\eqref{eq:BC}, though here, the $\mathcal{NN}$ takes two additional inputs $\mathbf{b}(t)$ and $\mathbf{c}$:
\begin{eqnarray}
  \mathbf{x}(t+\Delta t)&=&\mathcal{NN}_ \theta\big(\mathbf{x}(t),\mathbf{b}(t),\mathbf{c}\big), \label{eq:SIFC}\\
  \mathbf{x}(t-\Delta t)&=&\mathcal{NN}_ \phi\big(\mathbf{x}(t),\mathbf{b}(t),\mathbf{c}\big),  \label{eq:SIBC}     
\end{eqnarray}
where $\mathbf{b}(t)$ is the time-varying forcing top-of-atmosphere incident solar radiation and $\mathbf{c}$ includes the land--sea mask and surface geopotential field (topography) through channel-wise concatenation. Table~\ref{tab:variables} lists all of the variables used in $\mathbf{x}$, $\mathbf{b}$, and $\mathbf{c}$. We have trained independent, separate models for $\Delta t=6$ and $24$~h, all on the $1^{\circ}\times1^{\circ}$ horizontal resolution for computational feasibility. 

The deterministic forecasting and backcasting neural networks ($\mathcal{NN}_ \theta$ and $\mathcal{NN}_ \phi$) have identical architectures, based on the Pangu-Weather architecture~\cite{bi2023accurate}, hereafter referred to as the ``{Transformer}''. The full architecture description is provided in the Supplementary Materials. The sole difference between $\mathcal{NN}_ \theta$ and $\mathcal{NN}_ \phi$ is the reversed input--output pairing, $[\mathbf{x}(t) \to \mathbf{x}(t-\Delta t)]$ instead of $[\mathbf{x}(t) \to \mathbf{x}(t+\Delta t)]$, drawn from the same training dataset. Note that backcasting here should not be confused with the common meteorological terms ``hindcast'' or ``reforecast'', which refer to a \textit{forecast} initialized retrospectively for a historical date (see Fig.~\ref{fig1:era5-plasim}). Note that this \textit{autoregressive} backcasting is also different from the backward prediction in cBottle \cite{brenowitz2025climate}, a novel \textit{non-autoregressive} sampling emulator, which can produce a backward trajectory of a pre-defined length at once.

For training, each field is standardized by its mean and standard deviation computed from the training set. Both models are trained on single-time-step $\Delta t$ predictions; multi-step trajectories (rollouts) are generated autoregressively at inference by feeding each predicted state back in as input.

The Transformer minimizes a weighted mean absolute error (MAE) loss. Below, $x$ represents an ERA5 training sample and $\hat{x}$ denotes the AI model prediction; the subscripts $a$, $z$, and $s$ index upper-air variables, pressure levels, and surface variables, respectively (e.g., $x_{s}$ includes surface variables and $\hat{x}_{a,z}$ denotes predicted upper-air variables). All losses are latitude weighted, with
$\langle f \rangle_w \equiv \sum_{\varphi,\lambda} w(\varphi)\,f /
\sum_{\varphi,\lambda} w(\varphi)$ denoting the latitude-weighted spatial mean and $w(\varphi)=\cos\varphi$ (here, $\lambda$ and $\varphi$ are longitude and latitude, respectively). The MAE loss is:
\begin{equation}
  \mathcal{L}_{\mathrm{Transformer}}
  = \frac{1}{N_a N_z}\sum_{a,z}
    \bigl\langle|\hat{x}_{a,z} - x_{a,z}|\bigr\rangle_w
  + \frac{\alpha}{N_s}\sum_{s}
    \bigl\langle|\hat{x}_{s} - x_{s}|\bigr\rangle_w,
  \label{eq:loss_transformer}
\end{equation}
where $\alpha=1/4$ gives differential weighting between upper-air and surface variables. $N_z=17$, $N_a=5$, and $N_s=9$, are the number of vertical levels, upper-air variables, and surface variables, respectively.

The ERA5 forecasting and backcasting Transformers are trained with the AdamW optimizer and a OneCycleLR learning-rate schedule. We have explored a range of hyperparameters, e.g., the peak learning rate, learning-rate schedule, and number of epochs, and selected the best-performing configuration based on the one-time-step ($1\Delta t$) validation MAE. At the end, we have found the same hyperparameters for the best-performing forecasting and backcasting models, which have, overall, similar $1\Delta t$ accuracy, comparable to that of the official Pangu-Weather forecast, at the same $\Delta t$ (Fig.~\ref{fig:SI-one-step-RMSE}). The complete architectural and training hyperparameters are listed in Table~\ref{tab:hyperparams}.\\

\noindent\textit{Ensemble generation}\\
Ensembles are generated by adding perturbations to the initial conditions. Perlin noise following \cite{bi2023accurate} is added to all standardized fields of the initial condition ($T$, $u$, $v$, $q$, $Z$ on all pressure levels, and the surface variables). Each member is drawn from an independent random seed, and the perturbation amplitude is a function of a reference amplitude $\epsilon_0=0.1$. The initial conditions are taken once a month in 2020 and 2021, and a 16-member ensemble is generated for each. Each two-dimensional perturbation field is a 3-octave sum on the $180\times360$ grid with periods of (4, 8, 16) cycles across the domain (spatial scales of roughly $90^{\circ}$, $45^{\circ}$, and $22.5^{\circ}$ per cycle) and octave weights of (0.2, 0.1, 0.05). 

Hardware (GPU) precision and details can matter in small-amplitude perturbation experiments \cite{selz2026can}: apparent butterflies can be hardware artifacts, as the rapid DKE growth of several AIWP models has been shown to be GPU numerical noise that vanishes or shrinks on CPUs. Therefore, we have carefully ensured that all training of and inference with our AI models (for ERA5, PlaSim, and Lorenz 96) use single floating-point precision (FP32). We have disabled TF32 and conducted all experiments on the same cluster (CISL Derecho) and same GPUs (A100). We have also confirmed that setting $\epsilon_0 = 0$ leads to DKE $=0$ over time in Pangu-Weather used in Fig.~\ref{fig:downsampling}. To further see the effect of hardware, Fig.~\ref{fig:SI-sensitivity-Fig4-tf32-global} presents experiments with Pangu-Weather and lower precision (TF32), which show noticeably larger DKE growth. This is due to state-dependent random perturbations that are injected into the calculations over time due to lower precision, consistent with the results reported in \cite{selz2026can}. 

We have further verified that the DKE growth in Fig.~\ref{fig:downsampling}A is robust to the averaging domain (global versus $60^\circ$S--$60^\circ$N) and to the perturbation type (Perlin versus Gaussian); see Fig.~\ref{fig:SI-sensitivity-Fig4-tf32-global}. Fig.~\ref{fig:downsampling}A excludes latitudes poleward of $60^\circ$, where the freely running Pangu-Weather (with small $\Delta t$) can develop unphysical behaviors. Part of the rapid DKE growth of the $\Delta t = 1$~h model at high latitudes (included in total wavenumbers in Figs.~\ref{fig:downsampling}B and \ref{fig:SI-error-spectra}) reflects such instabilities rather than learned dynamics. Cleanly disentangling the two requires further study. \\

\noindent\textit{Forecasting-backcasting AI model: Preliminary exploration}\\
We have conducted preliminary exploration of building a single AI model that performs both forecasting and backcasting, as such a model might better learn dynamics. This bidirectional AI model can be represented as 
\begin{eqnarray}
  \mathbf{x}(t+a\Delta t)&=&\mathcal{NN}_ \vartheta\big(\mathbf{x}(t),a,\mathbf{b}(t),\mathbf{c}\big), \label{eq:SIFB}
\end{eqnarray}
where $a=+1$ (forecast) or $-1$ (backcast) determines the direction of the time evolution. This single neural network is trained with all training samples, $[\mathbf{x}(t) \to \mathbf{x}(t+\Delta t), a=+1]$ and $[\mathbf{x}(t) \to \mathbf{x}(t-\Delta t), a=-1]$, combined. The time indicator $a$ can be fed into the model in different ways. For the Transformer, we use a learned directional embedding, a vector of the token dimension added to the patch embedding of every token, as is done for the positional embedding. Apart from this embedding, the architecture, the weighted MAE loss of Eq.~(\ref{eq:loss_transformer}), and all training settings are the same as before (Table~\ref{tab:hyperparams}). We have only trained this model for $\Delta t = 24$~h. This preliminary investigation has not shown any noticeable improvement in accuracy or any change in DKE growth, though the design space for encoding $a$ remains largely unexplored. As described later, we reach the same conclusion with a preliminary exploration in PlaSim, for which we also tried a cyclic loss. 

\subsection{Numerical weather prediction (NWP) model: ICON}
As a physics-based NWP model reference, we use ICON, an operational global model from the German Weather Service and Max Planck Institute for Meteorology, which solves the discretized governing differential equations on an icosahedral grid. The ICON data (the DKE in Figs.~\ref{fig1:era5-plasim}C and~\ref{fig:downsampling}A) used in this study are provided by~\cite{selz2023can}, where the ICON ensembles are run at both 2.5~km and 20~km horizontal resolution.

\subsection{Atmosphere-only GCM: PlaSim}

\noindent\textit{GCM setup and data}\\
PlaSim~\cite{Fraedrich2005} is an intermediate-complexity global climate model (GCM) that couples a spectral atmospheric dynamical core, which solves the primitive equations for vorticity, divergence, temperature, and humidity, to a simplified land model. Sea surface temperature (SST) and sea ice are prescribed as boundary conditions that repeat identically each year. Numerical integrations are performed at T42 spectral truncation mapped onto a $64\times128$ Gaussian grid with 10 native vertical sigma levels. We have performed 110 years of integration. The first 60 years are discarded as spin-up (admittedly, unnecessarily too long). After integration, the 3D atmospheric state is interpolated onto 13 fixed pressure levels (Table~\ref{tab:variables}). Six-hourly data from simulation years 61--100 (40 years) are used for training and after a 4-year gap, year 105 is used for validation and checkpoint selection. Initial conditions taken every 50 days in Years 106--109 are employed for testing.

 We train separate, independent AIWP models with $\Delta t = 6$ and $24$~h for Transformer and $\Delta t = 24$~h for SFNO and Diffusion.  \\

\noindent\textit{Forecasting and backcasting AIWP models: Transformer, Neural Operator, and Diffusion}\\
For PlaSim, we train forecast and backcast AIWP models based on three architectures: the same Transformer used for ERA5, a spherical Fourier neural operator (SFNO)~\cite{bonev2023spherical}, and a diffusion transformer (Diffusion, hereafter)~\cite{peebles2023scalable}. Full architecture descriptions are provided in the Supplementary Materials. The same Eqs.~\eqref{eq:SIFC}-\eqref{eq:SIBC}, single-time-step training, autoregressive inference, and standardization described for ERA5 are used. 

Beyond the prognostic state $\mathbf{x}$, the Transformer and SFNO receive the time-varying forcings $\mathbf{b}$ and static fields $\mathbf{c}$ (Table~\ref{tab:variables}) through channel-wise concatenation, whereas the Diffusion uses them as cross-attention context and is additionally conditioned on day of year and hour of day. For the forecasting Diffusion model, Eq.~\eqref{eq:SIFC} represents a sample from the learned conditional distribution $p_{\theta}\!\left(\mathbf{x}(t+\Delta t)\mid\mathbf{x}(t),\mathbf{b}(t),\mathbf{c}\right)$ rather than a deterministic mapping to a unique future state. The same applies to the backcasting Diffusion model.

All three architectures minimize a latitude-weighted loss using the notation defined earlier for ERA5. The Transformer minimizes the weighted MAE loss of Eq.~\eqref{eq:loss_transformer}. The SFNO minimizes a latitude-weighted mean squared error (MSE) on the predicted state:
\begin{equation}
  \mathcal{L}_{\mathrm{SFNO}}
  = \frac{1}{N_a N_z + N_s}\left[
    \sum_{a,z}
    \bigl\langle(\hat{x}_{a,z} - x_{a,z})^{2}\bigr\rangle_w
  + \sum_{s}
    \bigl\langle(\hat{x}_{s} - x_{s})^{2}\bigr\rangle_w
  \right],
  \label{eq:loss_sfno}
\end{equation}
where atmospheric and surface fields share a single normalizer $N_a N_z + N_s$. $N_z=13$, $N_a=5$, and $N_s=4$, are the number of vertical levels, upper-air variables, and surface variables, respectively. The Diffusion AIWP model is trained by denoising score matching and minimizes the same latitude-weighted MSE, applied to the predicted noise $\hat{{\varepsilon}}$ :
\begin{equation}
  \mathcal{L}_{\mathrm{Diffusion}}
  = \frac{1}{N_a N_z + N_s}\left[
    \sum_{a,z}
    \bigl\langle(\hat{\varepsilon}_{a,z}
                - \varepsilon_{a,z})^{2}\bigr\rangle_w
  + \sum_{s}
    \bigl\langle(\hat{\varepsilon}_{s}
                - \varepsilon_{s})^{2}\bigr\rangle_w
  \right],
  \label{eq:loss_dit}
\end{equation}
where $\varepsilon$ is the noise injected at diffusion step $\tau_d \in \{1,\ldots,N_{\mathrm{diff}}\}$ ($\tau_d$ is the diffusion step index, distinct from the forecast lead time $\tau$), with $N_{\mathrm{diff}}=100$ and a cosine noise schedule.

All PlaSim models are trained with the AdamW optimizer; the learning-rate schedules and other training hyperparameters are listed in Table~\ref{tab:hyperparams}.
As for ERA5, we have explored a range of hyperparameters (e.g., peak learning rate, learning-rate schedule, and number of epochs). For Transformers, the forecasting and backcasting models are tuned independently based on their single-step ($1\Delta t$) validation MAE, which yields the same hyperparameters except for the peak learning rate (the $1\Delta t$ errors of the forecast and backcast models are overall comparable, especially for $\Delta t=24$~h; see Fig.~\ref{fig:SI-one-step-RMSE}). 
We also conducted hyperparameter optimization of the Transformer models against the error of the day-3 prediction, but this does not change any key results or conclusions, so all results reported in the paper are from models whose hyperparameters are chosen based on the $1\Delta t$ loss. For the SFNO and Diffusion models, hyperparameters are tuned on the forecasting model only, and the backcasting model adopts the same configuration. \\

\noindent\textit{Ensemble generation}\\
For all AIWP models of PlaSim, ensembles are generated by adding Perlin noise perturbations to initial conditions, following what we did for the ERA5 AIWP models. Because the Diffusion model is stochastic and can otherwise generate ensemble spread intrinsically through its sampling process, we run it in a ``quasi-deterministic'' mode following \cite{selz2026can}: the sampling seed is kept constant across ensemble members, so that ensemble spread arises solely from the explicit initial perturbations (with pre-defined amplitude) rather than from random sampling. Initial conditions are taken every 50 days during simulation years 106--109, and a 16-member ensemble is generated for each. Each two-dimensional noise field is constructed as for ERA5. 

For the numerical integration of the physics-based PlaSim model, we use the PlaSim's perturbation scheme implemented in ~\cite{ragone2018computation}. We generate a 16-member ensemble for each initial condition by adding random white noise of small amplitude ($10^{-8}$) to the spectral coefficients of the logarithmic surface pressure. 
Each perturbed state is advanced with the $\delta t=20$~min time step. 
The first 6~h is discarded as spin-up to allow rapid dynamical adjustment of the unphysical noise; at 6~h, we then subtract the ensemble mean to obtain the perturbations, which are scaled so that the initial ensemble spread (DKE) matches that of the corresponding AIWP models at 300~hPa (as shown in Fig.~\ref{fig1:era5-plasim}C). \\

\noindent\textit{Forecasting-backcasting AI model: Preliminary exploration}\\
As done for ERA5, we have explored a single bidirectional forecasting-backcasting Transformer for PlaSim, following Eq.~\eqref{eq:SIFB}. Here, we have additionally experimented with a cycle-consistency constraint in the loss function, which requires that one $\Delta t$ forward ($a=+1$) followed by one $\Delta t$ backward ($a=-1$) returns the original state. However, in these preliminary explorations, neither approach resulted in any improvement in accuracy or any change in the DKE growth.

\subsection{Canonical multi-scale, chaotic system: Two-scale Lorenz 96}
\noindent\textit{Equations and Data}\\
We use the two-scale Lorenz 96 model~\cite{lorenz1996predictability}, a canonical testbed for multi-scale, chaotic dynamics. Slow- and large-scale variables $X_i$ and fast- and small-scale variables $Y_{i,j}$ are governed by
\begin{align}
    \frac{dX_i}{dt} &= X_{i-1}(X_{i+1}-X_{i-2}) - X_i + F
                       - \frac{h\gamma}{\beta}\sum_{j=1}^J Y_{i,j},
    \label{eq:L96_X}\\
    \frac{dY_{i,j}}{dt} &= -\gamma\beta\,Y_{i,j+1}(Y_{i,j+2}-Y_{i,j-1})
                             - \gamma Y_{i,j} + \frac{h\gamma}{\beta}X_i,
    \label{eq:L96_Y}
\end{align}
with periodic boundary conditions ($i=1,\ldots,K$; $j=1,\ldots,J$). We adopt standard parameters~\cite{wilks2005effects} $K=8$, $J=32$, $F=20$, $h=1$, and $\beta=\gamma=10$, giving a 264-dimensional state vector and chaotic dynamics. Time in this system is commonly measured in model time units (MTU). Based on the error doubling time of the large scales, past studies \cite{lorenz1996predictability, thornes2017use} have estimated that with the current parameters, $1~\mathrm{MTU} \approx 5$~days, which we have adopted in all figures showing results from the Lorenz 96 system.

The forward dynamics is chaotic and dissipative. In this system, dissipation is stronger at the smaller scales by a factor of $\gamma=10$. 

To generate ground truth and training data, Eqs.~\eqref{eq:L96_X}-\eqref{eq:L96_Y} are integrated with a fourth-order Runge--Kutta scheme at $\delta t = 0.005$~MTU in double floating-point precision (FP64). The initial condition is drawn with independent standard normal values for all 264 components; after a 5{,}000-step (25~MTU) spin-up to discard transients, the subsequent $10^7$ steps are saved at every $\delta t$ in double precision.

The $10^7$-step dataset is sequentially partitioned into 90\% training and 10\% validation, and all variables are standardized using the training set's statistics. The validation segment is used for hyperparameter pruning, the learning-rate schedule, early stopping, and model selection (more details below). After model selection, an additional independent test trajectory equal in length to the validation segment is generated. The 100 evaluation initial conditions are drawn from this test trajectory and are spaced 1{,}000~$\delta t$ (5~MTU, more than 100 Lyapunov times) apart, far beyond the system's decorrelation time.

For the numerical results in Fig.~\ref{fig:rmse_hovmoller_1024}A, we numerically re-integrate the equations from a reference state on the archived trajectory both forward and backward in time, the latter using the same Runge--Kutta scheme with a negated time step $-\delta t$, in both single (FP32) and double (FP64) precision (the former is used to match the precision of AI models' training and inference). To prevent floating-point overflow, state values in the numerical integration are capped at $\pm10^{10}$; the backward trajectory reaches this cap within a fraction of an MTU. In Fig.~\ref{fig:rmse_hovmoller_1024}A, the saturated color regions correspond to $|X_i|$ exceeding the color-scale range as the backcast solution grows without bound; in Fig.~\ref{fig:dke_1024}, crosses mark the last lead time at which the ensemble DKE of the backcasting numerical integration remains below $10^3$ (the plotted range).\\

\noindent\textit{Forecasting and backcasting AI models: Multilayer perceptron}\\
Following Eqs.~\eqref{eq:FC} and~\eqref{eq:BC}, we train independent forecast and backcast AI models for the Lorenz 96 system. Each is a multilayer perceptron with ReLU activations, of width $W$ (neurons per hidden layer) and depth $D$ (number of hidden layers), that outputs the full state $\mathbf{x}$ at $t \pm \Delta t$ directly. The AI-model prediction interval is $\Delta t = n\,\delta t$ with $n \in \{1, 5, 10\}$ (reminder that $\delta t = 0.005$~MTU is the numerical
solver's time step). Note that the largest time step of AI models, $10\delta t = 0.05$~MTU~$=0.25$~day, is comparable to one Lyapunov time ($1/\lambda \approx 0.04$~MTU, from the measured leading finite-time Lyapunov exponent, FTLE, in Table~\ref{tab:si_lyapunov}). 

The Lorenz 96 results are based on 12 independently trained AI models:
\begin{itemize}
    \item Forecast and backcast models,
    \item Models with $\mathbf{x}=X_i$ and $\mathbf{x}=(X_i, Y_{i,j})$ (from concatenation),
    \item Models with $\Delta t=n \delta t$ where  $n=1, 5, \mathrm{or} \, 10$.
\end{itemize}

For each $n$, training pairs $\big(\mathbf{x}(t),\,\mathbf{x}(t \pm n\delta t)\big)$ use all saved samples and each model is trained on the full $9 \times 10^6$-sample training set. All models are trained with the Adam optimizer by minimizing the MSE of the predicted state.

We have selected the best performing AI models based on extensive hyperparameter optimization, though the key results and conclusions are robust with respect to these choices. We have explored a range of width $W \in \{512, 1024, 2048\}$ and depth $D \in \{3, 5, 7\}$ and four weight initialization seeds. Then for each fixed $(W,D)$ configuration and seed, the learning rate and batch size have been tuned independently for each of the twelve models with the Optuna framework~\cite{akiba2019optuna} and a Hyperband pruner~\cite{li2017hyperband} (Table~\ref{tab:l96_hyperparams}). All results are reported using the best-performing AI models with $W=1024$ and $D=7$, selected based on the criterion of using the same $(W,D)$ for forecast and backcast that produce similar lowest single-step validation error. Single time-step (one $\Delta t$) forecast and backcast RMSE are comparable for all models (backcast-to-forecast RMSE ratios of $0.90-1.13$ except for one case; see Table~\ref{tab:single_step_rmse}), indicating that the forecast--backcast asymmetry in Figs.~\ref{fig:rmse_hovmoller_1024} and~\ref{fig:dke_1024} is dominated by autoregressive rollout rather than by differences in the single time-step accuracy of the AI models.\\

\noindent\textit{Ensemble generation}\\
Ensembles are generated by adding random perturbations to
initial conditions: independent Gaussian noise of amplitude $\epsilon_0 = 0.1$ is added to all components of the initial state $\mathbf{x}$ to produce a 100-member ensemble. Reduced-amplitude ensembles are obtained by scaling down $\epsilon_0$.  \\

\noindent\textit{Computation of leading Lyapunov exponents in Lorenz 96}\\
In multi-scale chaotic systems like Lorenz 96, the leading global Lyapunov exponent is set by the fast, small-scale variables ($Y$): a generic infinitesimal perturbation asymptotically grows at this fastest rate, so the infinite-time measure is uninformative about the large scales ($X$). Predictability measures such as the error doubling time~\cite{thornes2017use} are informative about the large scales, but only because they track finite-amplitude errors, whose fast, small-scale components saturate while the large-scale error continues to grow. We instead require the growth rate of infinitesimal perturbations as they impact the large-scale ($X$) variables alone. Thus, for the numerical and AI models, we estimate a \textit{projected} FTLE: continuous perturbation rescaling keeps the perturbation infinitesimal, following the classical algorithm of Benettin~\textit{et al.}~\cite{benettin1980lyapunov} as implemented by Wolf \textit{et al.}~\cite{wolf1985determining}, and the growth rate is measured, over a finite window, from the $X$ components alone. Because the estimator requires only evaluations of the model itself (no tangent-linear equations), it is applied identically to the numerical and AI models. 

A random initial state $\mathbf{x}_0$ and a perturbed state $\mathbf{x}'_0 = \mathbf{x}_0 + \delta_0 \mathbf{v}_0$ (where $\mathbf{v}_0$ is a random unit vector in the full state space and $\delta_0=10^{-4}$) are integrated through the AI model (time step $\Delta t$) or the numerical model (time step $\delta t$). At each step $m$, the full perturbation vector $\Delta \mathbf{x}_m = \mathbf{x}'_m - \mathbf{x}_m$ is recorded, and the perturbed state is re-normalized to magnitude $\delta_0$ along the new divergence vector $\mathbf{v}_m = \Delta \mathbf{x}_m / \|\Delta \mathbf{x}_m\|$. The projected forward FTLE of the large-scale variables, $\lambda^{(X)}$, is the time average over a finite window of $N_{\mathrm{step}}$ steps:
\begin{equation}
    \lambda^{(X)} = \frac{1}{N_{\mathrm{step}} \Delta t} \sum_{m=1}^{N_{\mathrm{step}}}    \ln\left( \frac{\|\Delta \mathbf{x}^{(X)}_m\|}{\delta_0 \|\mathbf{v}^{(X)}_{m-1}\|} \right),
\end{equation}
where, to isolate the predictability of the large scales, the logarithmic growth rate uses exclusively the norm of the $X$ components. The fast-variable exponents $\lambda^{(Y)}$ in Table~\ref{tab:si_lyapunov} are computed similarly but using the $Y$ components. The averaging window spans $0.1$--$0.3$~MTU after a $0.1$-MTU transient, so $N_{\mathrm{step}} = 0.2~\mathrm{MTU}/\Delta t$ (40 steps for the numerical model and the $1\delta t$ AI models; 4 steps at $\Delta t = 10\delta t$).

To capture the variability of predictability across the reference trajectory, reported values represent the mean and standard deviation evaluated over 20 independent initial conditions drawn from the reference trajectory.

The $\lambda^{(X)}$ and $\lambda^{(Y)}$ for the ground truth and some of the key AI models are reported in Table~\ref{tab:si_lyapunov}. \\

\subsection{Evaluation metrics}

Forecast and backcast skills are assessed using two complementary metrics: the anomaly correlation coefficient (ACC) and the root mean square error (RMSE), which measure the accuracy of trajectory predictions. We also compute the difference kinetic energy (DKE), which quantifies the growth of initial perturbations and characterizes the chaotic behavior of the system. ACC and RMSE are evaluated for several predicted variables, whereas DKE is evaluated from the horizontal wind components. For Lorenz 96, RMSE and DKE are both based only on the large-scale variable, $X$.

Let $\tau$ denote forecast or backcast lead time, $i = 1,\ldots,N$ the grid point index, and $w_{i} = \cos\varphi_{i}$ the area weight for latitude $\varphi_{i}$. For deterministic verification, $\hat{x}_{i}(\tau)$ and $x_{i}(\tau)$ are the predicted and ground truth values, and $\bar{x}_{i}^{\,\mathrm{c}}$ is the climatological mean at grid point $i$. For ensemble verification, $u_{i}^{(m)}(\tau)$ and $v_{i}^{(m)}(\tau)$ are the zonal and meridional winds of member $m = 1,\ldots,M$, with instantaneous ensemble means $\tilde{u}_{i}(\tau)$ and $\tilde{v}_{i}(\tau)$. All metrics are averaged over the evaluation cases (inference from many initial conditions in the test set) for each system to yield lead-time-dependent skill curves.\\

\noindent\textit{Anomaly correlation coefficient (ACC)}\\
ACC measures the area-weighted pattern similarity between predicted and target anomalies relative to climatology (defined as the calendar-day mean over 1979--2018 for ERA5 and years 61--100 for PlaSim),
\begin{equation}
\mathrm{ACC}(\tau)
  = \frac{\displaystyle\sum_{i=1}^{N}
          w_{i}
          \bigl(\hat{x}_{i}(\tau) - \bar{x}_{i}^{\,\mathrm{c}}\bigr)
          \bigl(x_{i}(\tau)       - \bar{x}_{i}^{\,\mathrm{c}}\bigr)}
         {\sqrt{\displaystyle\sum_{i=1}^{N}
                w_{i}\,\bigl(\hat{x}_{i}(\tau) -
                \bar{x}_{i}^{\,\mathrm{c}}\bigr)^{2}}\;
          \sqrt{\displaystyle\sum_{i=1}^{N}
                w_{i}\,\bigl(x_{i}(\tau) -
                \bar{x}_{i}^{\,\mathrm{c}}\bigr)^{2}}}.
\label{eq:acc}
\end{equation}
Note that ACC primarily measures the large-scale forecast accuracy.
The conventional skill threshold for weather prediction is $\mathrm{ACC} > 0.6$~\cite{krishnamurti2003improved}. For all ERA5 and PlaSim results, we define the asymmetry as the ratio of forecast to backcast lead times at $\mathrm{ACC} = 0.6$.\\

\noindent\textit{Root mean square error (RMSE)}\\
RMSE quantifies prediction error as
\begin{equation}
\mathrm{RMSE}(\tau)
  = \sqrt{\frac{\displaystyle\sum_{i=1}^{N}
                w_{i}\,\bigl(\hat{x}_{i}(\tau) - x_{i}(\tau)\bigr)^{2}}
               {\displaystyle\sum_{i=1}^{N} w_{i}}}.
\label{eq:rmse}
\end{equation}
\\
For Lorenz 96, RMSE is defined as in Eq.~\eqref{eq:rmse} with uniform weights over the
$K=8$ large, slow variables $X$. The asymmetry in Lorenz 96 is then defined as the backcast/forecast RMSE ratio at a fixed lead (1.5 days; Fig.~\ref{fig:rmse_hovmoller_1024}).\\

\noindent\textit{Difference kinetic energy (DKE)}\\
DKE measures ensemble spread rather than error against a target. It is the area-weighted mean of half the ensemble variance of the horizontal wind components, computed about the instantaneous ensemble mean ($\tilde{u}_{i}(\tau)$, $\tilde{v}_{i}(\tau)$), distinct from the calendar-day climatological mean $\bar{x}_{i}^{\,\mathrm{c}}$ used in ACC. Analogous to the kinetic component of the difference total energy of \cite{zhang2003effects}, it quantifies spread directly in dynamical units:
\begin{equation}
\mathrm{DKE}(\tau)
  = \frac{\displaystyle\sum_{i=1}^{N} w_{i}\,
          \tfrac{1}{2}\bigl[\sigma_{u,i}^{2}(\tau) + \sigma_{v,i}^{2}(\tau)\bigr]}
         {\displaystyle\sum_{i=1}^{N} w_{i}},
\qquad
\sigma_{u,i}^{2}(\tau)
  = \frac{1}{M-1}\sum_{m=1}^{M}
    \bigl(u_{i}^{(m)}(\tau) - \tilde{u}_{i}(\tau)\bigr)^{2},
\label{eq:dke}
\end{equation}
where $\sigma_{u,i}^{2}$ and $\sigma_{v,i}^{2}$ are the unbiased ensemble variances of the two horizontal wind components at grid point $i$, $w_{i}$ is the area weight, $M$ is the ensemble size, and $N$ the number of grid points. Units are $\mathrm{m}^{2}\,\mathrm{s}^{-2}$. Globally averaged DKE at 300~hPa is reported, unless specified otherwise.

For the Lorenz 96 system, DKE is analogously defined as half the sum of the ensemble variances of the large-scale variables:
\begin{equation}
    \mathrm{DKE}_{\mathrm{Lorenz \, 96}}(\tau) = \frac{1}{2} \sum_{i=1}^K \frac{1}{M-1} \sum_{m=1}^M \bigl(X_i^{(m)}(\tau) - \tilde{X}_i(\tau)\bigr)^2,
\label{eq:dke_l96}
\end{equation}
where $X_i^{(m)}$ is the large-scale state of ensemble member $m$ and $\tilde{X}_i$ is the instantaneous ensemble mean.

\newpage

\section{Supplementary Materials}

\subsection{Time reversal in multi-scale, dissipative, chaotic systems and the $2^\mathrm{nd}$ law of thermodynamics}
This section further describes why predicting the past of a multi-scale, dissipative, chaotic system, such as the atmosphere, should be practically impossible. It then explains why backcasting AI models can nevertheless be skillful without violating any of these arguments, or the second law of thermodynamics. We use three examples of increasing relevance: the heat equation, Lorenz 63, and the two-scale Lorenz 96 system (Methods and Data). \\

\noindent\textit{{Multi-scale dynamics: Heat equation and Hadamard ill-posedness}} \\
The linear, one-dimensional heat equation, $\partial T/\partial t=\kappa\,\partial^2T/\partial x^2$ with diffusivity $\kappa>0$ and spatial coordinate $x$, evolves each Fourier mode of wavenumber $k$ independently:
\begin{equation}
\hat{T}(k,t)=\hat{T}(k,0)\,e^{-\kappa k^{2}t}.
\label{eq:heat}
\end{equation}
Forward in time ($t>0$), perturbations decay, most quickly at the smallest spatial scales; the analytical and numerical solutions are stable to initial-condition errors. Reversing time ($t^*=-t>0$) flips the exponent: a perturbation of amplitude $\varepsilon$ at wavenumber $k$, unavoidable in any real computation or measurement, grows backward as $\varepsilon\,e^{+\kappa k^{2}t^*}$ and reaches order unity by ${t}^*_e(k)=\frac{\ln(1/\varepsilon)}{\kappa k^{2}}$. Even at double floating-point precision ($\varepsilon\sim10^{-16}$, $\ln(1/\varepsilon)\approx37$), ${t}^*_e$ quickly collapses quadratically with wavenumber: the smallest scales dominate the backcasting solution first, and in a system with a broad range of scales, they do so almost immediately. This is the classical statement that the backward heat problem is ill-posed in the sense of Hadamard \cite{hadamard2014lectures,payne1975improperly}. Note that here, the exponents in forecasting and backcasting have the same amplitudes (as a function of length scale) but opposite signs ($-\kappa k^{2}$ versus $+\kappa k^{2}$). \\

\noindent\textit{{Chaotic dynamics: Lorenz 63}} \\
Chaotic dissipative systems further refine the picture of backcasting. The nonlinear, single-scale  Lorenz 63 system \cite{lorenz1963}, 
\begin{align}
\frac{dx}{dt} &= \sigma(y-x),\label{eq:L63x} \\
\frac{dy}{dt} &= x(\rho-z)-y,\label{eq:L63y} \\
\frac{dz}{dt} &= xy-\beta z, \label{eq:L63z}
\end{align}
with standard parameters ($\sigma=10$, $\rho=28$, $\beta=8/3$) has Lyapunov exponents $(\lambda_{1},\lambda_{2},\lambda_{3})\approx(0.91,\,0,\,-14.57)$ \cite{viswanath1998lyapunov}. Their sum equals the phase-space divergence:
\begin{equation}
 \sum_{i=1}^{3} \lambda_i = -(1+\sigma+\beta) \approx-13.67, 
\end{equation}
i.e., \textit{volumes contract}, and trajectories settle onto an \textit{attractor}, the Lorenz strange attractor, while nearby states separate exponentially at the modest rate $\lambda_{1}\approx +0.91$ (Movie~S1). Reversing time negates the Lyapunov spectrum, which reorders to $\approx(+14.57,\,0,\,-0.91)$; see below for more discussion on this reordering. 

Two major consequences follow, and Movie S1 shows both. First, the reversed system is dramatically more unstable along trajectories: infinitesimal errors $e$-fold $14.57/0.91 \approx 16$ times faster in backcasting compared with forecasting, so an initial error reaches order one $16\times$ faster. Second, and more damning, the sum of the reversed exponents is 
\begin{equation}
-\sum_{i=1}^{3} \lambda_i = +(1+\sigma+\beta) \approx +13.67. 
\end{equation}
Backward in time, phase-space volumes \textit{expand}, and the \textit{attractor} of the forward dynamics is a \textit{repeller} of the backward dynamics. Any perturbation transverse to the attractor, which forward dynamics would harmlessly contract away, is exponentially amplified backward at rates up to $e^{+14.57 t^*}$, and trajectories are expelled from the attractor altogether (see Movie~S1). 

As a result, numerical backcasting of \textit{single-scale} Lorenz~63 fails for two distinct reasons: errors grow faster and into regions the dynamics never visits. Note that here, unlike for the heat equation, time-reversal not only negates the signs of exponents but also changes the amplitude of the leading exponent, leading to $16\times$ faster growth of errors in backcasting. These two failures follow from two distinct properties of Lorenz 63's Lyapunov spectrum under time reversal: A)~The sign reversal of all exponents, and B) $|\lambda_3|>\lambda_1$. Next, we discuss the conditions under which (A) and (B) apply to more complex systems such as the atmosphere.

As for (A), if the dynamics is invertible, admits an ergodic invariant probability measure, and satisfies a mild integrability condition on the tangent dynamics, the Oseledets multiplicative ergodic theorem guarantees a well-defined Lyapunov spectrum, and the exponents of the time-reversed dynamics are the forward exponents with reversed signs~\cite{oseledets1968multiplicative,young2013mathematical,ginelli2013covariant,hoover1988negative}. For dissipative systems such as the atmosphere \cite{ghil2020physics}, the invariant measure is supported on the attractor, so the reversal applies to trajectories on the attractor itself \cite{eckmann1985ergodic}; e.g., the initial conditions that we choose from forward numerical integrations of Lorenz 96 and PlaSim, or from ERA5 reanalysis. Off-attractor states diverge under the reversed flow and have no backward Lyapunov exponents~\cite{eckmann1985ergodic}; such states are not available from ERA5 reanalysis or any forward numerical integration (Lorenz 96 or PlaSim) and are not of concern here. Therefore, (A) is broadly true.  

Let us order the $L$ Lyapunov exponents of a system from the most positive  $\lambda_1$ (leading) to the most negative $\lambda_L$. Then, property (B) is that $|\lambda_L| > \lambda_1$, so that upon reversal and reordering, the leading backward exponent exceeds the leading forward one, making backward error growth exponentially faster than forward. In three-dimensional dissipative, chaotic systems such as Lorenz~63 this is automatic: the exponent sum equals the (negative) phase-space divergence and the intermediate exponent vanishes, so $|\lambda_3| = \lambda_1 + (\sigma+1+\beta) > \lambda_1$ (equivalently, the Kaplan--Yorke dimension lies below three~\cite{kaplan1979chaotic}). In higher dimensions, however, a negative exponent sum only guarantees that the contracting directions collectively outweigh the expanding ones, so $|\lambda_L| > \lambda_1$ is not guaranteed in general. But the picture changes once we consider multi-scale systems. In such systems, dissipation acts most strongly on the smallest and fastest scales, leading to $|\lambda_L| > \lambda_1$, as documented in computed Lyapunov spectra of Lorenz~96 \cite{karimi2010extensive,carlu2019lyapunov} and of atmospheric and coupled climate models~\cite{decruz2018exploring,vannitsem2017predictability}.
Therefore, both (A) and (B) are expected to apply to the atmosphere. 

Considering the above discussion on the heat equation and Lorenz 63, backcasting of multi-scale, dissipative, chaotic systems should be practically impossible. Next, we discuss a canonical system that embodies all three characteristics, the two-scale Lorenz 96. \\ 

\noindent\textit{{Multi-scale, chaotic dynamics: Two-scale Lorenz 96}} \\
This system (Eqs.~\eqref{eq:L96_X}-\eqref{eq:L96_Y}) adds the ingredient the atmosphere has and Lorenz 63 lacks: multi-scale dynamics, specifically, faster dissipation at smaller scales $Y_{i,j}$. Under time reversal, the damping terms $-X_i$ and $-\gamma Y_{i,j}$ become anti-dissipative, with the fast variables, growing $\gamma=10$ times more strongly, blowing up first and dragging the large scales $X_i$ with them through the coupling. Consequently, backcasting via numerical integration diverges rapidly at both single and double floating-point precision (Fig.~\ref{fig:rmse_hovmoller_1024}A). \\

\noindent\textit{{The second law of thermodynamics, the arrow of time, and the asymmetry}}\\
It may seem that a skillful AI backcasting conflicts with the second law of thermodynamics: entropy increase selects a direction (arrow) of time, and the atmosphere, a forced-dissipative system full of mixing and irreversible diabatic processes, plainly should follow that arrow. However, coarse-graining removes the spatio-temporal scales at which the strongest dissipation happens, leading to a training set that is closer to reversible. Seen in a different way, coarse-graining mimics the classical remedy to Hadamard ill-posedness: regularization by truncating/damping the high wavenumbers, after which the time-reversed integrations become stable over a finite window \cite{lattes1969method,tikhonov1977solutions}. Furthermore, the training data contain only states on (or extremely near) the attractor, so the AI model never learns, and is never asked to represent, the explosive off-attractor directions (and as described above, because of the coarse-graining, the on-attractor instability the training data inherits is milder as well). A learned backcast does not integrate the reversed vector field; it estimates, by regression, the attractor state $\Delta t$ earlier that is most compatible with the current coarse state. It operates on the attractor, at the ``resolved'' scales, or not at all. 

In the end, the skillful AI backcasting does not violate the second law of thermodynamics; it just operates on a dataset that is closer to reversible due to coarse-graining and on-attractor sampling. The forecast-backcast asymmetry that we see in the hierarchy (Figs.~\ref{fig1:era5-plasim}B and \ref{fig:rmse_hovmoller_1024}B-C) is then a measure of the residual irreversibility (dynamical dissipation and, for the atmosphere, thermodynamic entropy production) that survives coarse-graining and is retained in the training data. This interpretation makes a testable prediction: the asymmetry should be largest where irreversible processes are strongest in the resolved dynamics of the atmosphere. The larger asymmetry we find in the tropics, where diabatic heating and convective dissipation dominate \cite{peixoto1991entropy,pauluis2002entropy,emanuel1994large}, compared with the extratropics (Table~\ref{tab:asymmetry}) is consistent with this prediction, and warrants further investigation.   

Finally, it is worth mentioning the double role ``coarse-graining'' might appear to play, and the distinction between the two operations this term refers to can be instructive. Microscopic dynamics (at the molecular level) are time-reversal symmetric; irreversibility and entropy growth emerge at the level of \textit{coarse-grained, macroscopic descriptions}\footnote{It is worth mentioning that Seif~\textit{et al.}~\cite{seif2021machine} has addressed the question of whether neural networks can learn the ``thermodynamic arrow of time'' from movies of microscopic processes.}, when information about the discarded degrees of freedom is given up, as formalized from Boltzmann's molecular chaos assumption to the projection-operator formalisms of Mori and Zwanzig \cite{zwanzig1960ensemble, mori1965transport,zwanzig2001nonequilibrium}. \textit{Thus, coarse-graining from molecular (microscopic) to continuous (macroscopic) dynamics is the operation through which the latter acquires its arrow of time and time-irreversibility} \cite{lebowitz1993boltzmann,zeh2007physical}. In the context of the work here, coarse-graining refers to the removal of the fast and small scales of the \textit{macroscopic dynamics}, which brings the training data \textit{closer to time-reversible}.  \\

\noindent\textit{{The role of stochasticity and the Mori-Zwanzig Formalism}}\\
All of the arguments above concern deterministic dynamics. Stochastic dynamics behave qualitatively differently under time reversal. Consider the linear ordinary differential equation $dx/dt=-a\,x$ with $a>0$, which, for example, governs the Fourier mode of the heat equation with $a=\kappa k^{2}$. The forecast solution is $x(t) \sim e^{-at}$, which exponentially decays. The time-reversed equation is $dx/dt^*=+a\,x$, where $t^*=-t>0$. Thus, the backcast solution is $x(t^*) \sim e^{+a t^*}$, which exponentially blows up. 

Now let's consider the stochastic version driven by Gaussian white noise $\eta(t)$:
\begin{equation}
\frac{dx}{dt} = -a\,x + \eta(t),
\label{eq:ou}
\end{equation}
i.e., an Ornstein--Uhlenbeck process \cite{gardiner2009stochastic}. When the noise is additive and the process is statistically stationary,  the equation obeyed by the time-reversed process \cite{anderson1982reverse, haussmann1986time} is:
\begin{equation}
\frac{d{x}}{dt^*} = -a\,{x} + \tilde{\eta}(t^*),
\label{eq:ourev}
\end{equation}
where $\tilde{\eta}$ is another realization of the same white noise. This is basically the same equation as \eqref{eq:ou}, showing that the stationary stochastic process is time-reversible. \textit{The past is exactly as predictable as the future, which is dramatically different from the behavior of the same equation when the noise vanishes} ($\eta=0$). 

The Mori--Zwanzig formalism \cite{mori1965transport,zwanzig1960ensemble,zwanzig2001nonequilibrium} reconciles the coarse-graining versus stochasticity explanations. By the Mori--Zwanzig formalism, eliminating the fast, small scales of a deterministic multiscale system  (coarse-graining) leaves resolved-scale dynamics that contain memory and a stochastic forcing \cite{chorin2000optimal}. This observation underlies stochastic climate modeling \cite{majda2001mathematical, christian2015stochastic, falasca2026physics} and stochastic parameterizations \cite{wouters2013multi,kondrashov2015data}. Coarse-grained deterministic data are thus statistically of the same kind as realizations of a stochastic process: the stochastic-process view of ERA5 and the coarse-graining explanation of the main text are one interpretation expressed in two languages.

\subsection{More details on neural network architectures}

\noindent\textit{The Transformer model}\\
The Transformer follows the three-dimensional Earth-specific transformer architecture of
\cite{bi2023accurate}. A 3D patch embedding groups every $2{\times}2{\times}2$ block in (level, latitude,
longitude) into a $240$-dimensional token.
These tokens pass through a hierarchical encoder--decoder of Swin-style shifted-window self-attention blocks in four stages of depth 2, 6, 6, and 2, each attending within a local window spanning two pressure levels, with 6, 12, 12, and 6 attention heads per stage.
The encoder halves the horizontal resolution at each stage and the decoder restores it; a sub-pixel deconvolution head and a recovery head reconstruct the full-resolution prognostic state. The ERA5 Transformer uses a window size of $2{\times}6{\times}10$. The PlaSim Transformer is identical except for a window size of $2{\times}6{\times}12$, matching the T42 grid; see Table~\ref{tab:hyperparams} for details. \\

\noindent\textit{{The spherical Fourier Neural Operator (SFNO) model}}\\
The SFNO~\cite{bonev2023spherical} uses spherical
harmonic transform (SHT), so that spectral filtering respects the Earth's spherical geometry. In plain terms, the model learns transformations in spherical spectral space rather than directly
on the latitude–longitude grid (used in FourCastNet \cite{pathak2022fourcastnet}). A point-wise encoder lifts the input to a $384$-dimensional channel space.
Twelve spectral blocks each apply a forward SHT, a learned degree-dependent linear
filter, and an inverse SHT, followed by a two-layer GELU multilayer perceptron (expansion ratio 2) with a residual connection and per-channel instance normalization.
A global residual connection links encoder to decoder. \\

\noindent\textit{The Diffusion model} \\
This is a conditional score-based diffusion model. This generative model learns
to reverse an artificial noising process and generates a target state through iterative denoising. A forward process adds Gaussian noise to the target state over $N_{\mathrm{diff}}{=}100$
steps with a cosine noise schedule. 
The denoiser is a vision transformer that tokenizes the input via a $2{\times}2$ patch
embedding into $1024$-dimensional tokens, to which a learned spherical-harmonics positional embedding is added.
Twelve self-attention blocks (16 heads each) operate under adaptive layer normalization
conditioned on the diffusion noise level and calendar position (day-of-year and
hour-of-day), setting per-block shift, scale, and gate parameters.
After each of the first 6 blocks, a cross-attention layer allows state tokens to attend
to boundary-forcing context tokens.
A noise-level-modulated unpatchify head projects back to the prognostic channels.

\providecommand{\checkmark}{\ensuremath{\surd}}

\newpage
\begin{table}[ht]
\centering
\caption{\textbf{Regional variation of the forecast-backcast asymmetry}. As defined in the main text, the asymmetry is the ratio of the lead times at which the forecast and backcast ACC drops to 0.6. Asymmetry~$>1$ means that backcasting is less skillful than forecasting. Asymmetry values are reported globally and separately for the extratropical Northern Hemisphere (NH: 30$^\circ$N--70$^\circ$N), Tropics: 30$^\circ$S--30$^\circ$N, and extratropical Southern Hemisphere (SH: 70$^\circ$S--30$^\circ$S) for both ERA5 and PlaSim and for a variety of architectures and $\Delta t$. For each variable and AI model, the largest value among the three regions is shown in bold. For ERA5, consistent across both $\Delta t$ values and variables at 3 vertical levels, the asymmetry is noticeably largest in the tropics. For PlaSim, the same trend is seen for Z500 and U250, but the trend is not robust for U850.}
\label{tab:asymmetry}
\begin{tabular}{ll|cc|cccc}
\hline
\multicolumn{2}{l}{ } & \multicolumn{2}{c}{\textbf{ERA5}} & \multicolumn{4}{c}{\textbf{PlaSim}} \\
\cmidrule(lr){3-4} \cmidrule(lr){5-8}
\multicolumn{2}{l}{\textbf{Architecture}} & \multicolumn{2}{c}{Transformer} & \multicolumn{2}{c}{Transformer} & SFNO & Diffusion \\
\cmidrule(lr){3-4} \cmidrule(lr){5-6} \cmidrule(lr){7-7} \cmidrule(lr){8-8}
\textbf{Variable} & \textbf{Region} & 24\,h & 6\,h & 24\,h & 6\,h & 24\,h & 24\,h \\
\hline
\multirow{4}{*}{U250}
 & Global  & 1.50 & 1.81 & 1.78 & 1.98 & 1.77 & 2.03 \\
 & NH      & 1.36 & 1.66 & 1.68 & 1.84 & 1.68 & 1.94 \\
 & Tropics & \textbf{1.91} & \textbf{2.23} & \textbf{2.04} & \textbf{2.15} & \textbf{2.00} & \textbf{2.32} \\
 & SH      & 1.42 & 1.74 & 1.72 & 1.87 & 1.79 & 1.99 \\
\hline
\multirow{4}{*}{Z500}
 & Global  & 1.40 & 1.63 & 1.63 & 1.76 & 1.59 & 1.94 \\
 & NH      & 1.35 & 1.64 & 1.59 & 1.73 & 1.48 & 1.89 \\
 & Tropics & \textbf{1.53} & \textbf{1.97} & \textbf{2.03} & \textbf{2.75} & \textbf{1.82} & \textbf{2.17} \\
 & SH      & 1.39 & 1.65 & 1.63 & 1.72 & 1.70 & 1.94 \\
\hline
\multirow{4}{*}{U850}
 & Global  & 1.47 & 2.03 & 1.82 & 2.35 & 1.77 & 2.19 \\
 & NH      & 1.38 & 2.07 & \textbf{1.80} & \textbf{2.34} & 1.75 & 2.22 \\
 & Tropics & \textbf{1.79} & \textbf{2.52} & 1.79 & 2.25 & 1.71 & \textbf{2.27} \\
 & SH      & 1.33 & 1.84 & \textbf{1.80} & 2.24 & \textbf{1.83} & 2.06 \\
\hline
\end{tabular}
\end{table}

\newpage
\begin{table}[ht]
\centering\caption{\textbf{Leading Lyapunov exponents of Lorenz 96 numerical and AI models.} The leading finite-time Lyapunov exponents (FTLE) $\lambda$ are estimated using the rescaling method (see Methods and Data).  The reported values are the mean and standard deviation over 20 independent initial conditions on the reference trajectory, each with an independent random initial perturbation of amplitude $\delta_0 = 10^{-4}$, computed over the window $0.1$--$0.3$~MTU (after discarding a $0.1$-MTU transient). The FTLE values are projected onto the slow ($X$) or fast ($Y$) variables.}\label{tab:si_lyapunov}\begin{tabular}{llc}\hline\textbf{System / AI model} & \textbf{Projection} &\textbf{$\lambda$ (MTU$^{-1}$)} \\\hline Ground truth                     & $X$ & 21.9 $\pm$ 6.3 \\Ground truth                     & $Y$ & 24.3 $\pm$ 4.0 \\$XY$, $1\delta t$ AI model        & $X$ & 26.8 $\pm$ 6.5 \\$XY$, $1\delta t$ AI model       & $Y$ & 26.8 $\pm$ 5.9 \\$X$, $1\delta t$ AI model   & $X$ & 3.05 $\pm$ 1.84 \\$X$, $10\delta t$ AI model  & $X$ & 2.81 $\pm$ 1.59 \\\hline\end{tabular}\end{table}

\begin{table}[htbp]
\centering
\caption{\textbf{Variable inventories for the Pangu-Weather, ERA5, and PlaSim AIWP models.}
  Upper-air prognostic state $\mathbf{x}_{\mathrm{ua}}$, surface prognostic state
  $\mathbf{x}_{\mathrm{sfc}}$, time-varying boundary forcings $\mathbf{b}$, and
  static fields $\mathbf{c}$ are listed.
  Checkmarks indicate inclusion; dashes indicate not applicable. The Pangu-Weather column
  describes the pretrained model on its native $0.25^{\circ}$ grid, which we use for inference only.
  The ERA5 Transformer column describes the data for the $1^{\circ}$ ERA5 Transformer. The PlaSim AIWP column describes data for the Transformer, SFNO, and Diffusion AIWP models. $^\dagger$We have trained versions of the PlaSim Transformer with sea-surface temperature as a prognostic variable (in $\mathbf{x}$) and the key observations and conclusions from Fig.~\ref{fig1:era5-plasim}B-C remain the same.}
\label{tab:variables}
\resizebox{\textwidth}{!}{%
\begin{tabular}{lllll}
\hline
\textbf{Category} & \textbf{Variable} & \textbf{Pangu-Weather}~\cite{bi2023accurate}  & \textbf{ERA5 Transformer} & \textbf{PlaSim AIWP} \\
\hline
                                        & Temperature ($T$)        & 13 levels & 17 levels & 13 levels \\
                                        & Zonal wind ($u$)         & 13 levels & 17 levels & 13 levels \\
Upper-air $\mathbf{x}_{\mathrm{ua}}$    & Meridional wind ($v$)    & 13 levels & 17 levels & 13 levels \\
                                        & Specific humidity ($q$)  & 13 levels & 17 levels & 13 levels \\
                                        & Geopotential ($Z$)& 13 levels & 17 levels & 13 levels \\
\hline
                                        & 2-m temperature                  & $\checkmark$ & $\checkmark$ & $\checkmark$ \\
                                        & 10-m zonal \& meridional wind    & $\checkmark$ & $\checkmark$ & ---          \\
                                        & Mean sea-level pressure          & $\checkmark$ & $\checkmark$ & ---          \\
Surface $\mathbf{x}_{\mathrm{sfc}}$     & Surface pressure                 & ---          & $\checkmark$ & $\checkmark$ \\
                                        & Skin temperature                 & ---          & $\checkmark$ & $\checkmark$ \\
                                        & 0--7\,cm soil water \& soil temp.& ---          & $\checkmark$ & ---          \\
                                        & Soil moisture                    & ---          & ---          & $\checkmark$ \\
                                        & Sea-surface temperature          & ---          & $\checkmark$ & ---          \\
\hline
Time-varying                            & top-of-atmosphere incident solar radiation & --- & $\checkmark$ & $\checkmark$ \\
forcings $\mathbf{b}$                   & Sea-surface temperature          & ---          & ---          & $\checkmark^\dagger$ \\
                                        & Sea-ice concentration            & ---          & ---          & $\checkmark$ \\
\hline
Static fields $\mathbf{c}$              & Land--sea mask                   & ---          & $\checkmark$ & $\checkmark$ \\
                                        & Surface geopotential      & ---          & $\checkmark$ & $\checkmark$ \\
\hline
   & Horizontal grid & $721\times1440$ ($0.25^{\circ}$) & $180\times360$ ($1^{\circ}$) & $64\times128$ (T42) \\
   & Pressure levels (hPa)
     & \begin{tabular}[t]{@{}l@{}}50, 100, 150, 200, 250,\\ 300, 400, 500, 600, 700,\\ 850, 925, 1000\end{tabular}
     & \begin{tabular}[t]{@{}l@{}}5, 10, 20, 30, 50, 70,\\ 100, 150, 250, 300, 400,\\ 500, 600, 700, 850,\\ 925, 1000\end{tabular}
     & \begin{tabular}[t]{@{}l@{}}50, 100, 150, 200, 250,\\ 300, 400, 500, 600, 700,\\ 850, 925, 1000\end{tabular} \\
   & Time interval $\Delta t$ & 1\,h, 3\,h, 6\,h, 24\,h & 6\,h, 24\,h & 6\,h, 24\,h \\
   & Training period   & 1979--2017 (39\,years) & 1979--2018 (40\,years) & 61--100 (40\,years) \\
   & Validation period & 2019 (1\,year)         & 2019 (1\,year)         & 105 (1\,year) \\
   & Test period       & 2020-2021 (2\,years)         & 2020-2021 (2\,years)         & 106--109 (4\,years) \\
\hline
\end{tabular}%
}
\end{table}

\newpage

\begin{table}[htbp]
\centering
\caption{\textbf{Architectural and training hyperparameters for ERA5 and PlaSim AIWP models.} Entries marked with a dash are not applicable. $^{\dagger}$\,Forecast\,/\,backcast peak learning rates.}
\label{tab:hyperparams}
\resizebox{\textwidth}{!}{%
\begin{tabular}{llcccc}
\hline
 & & \textbf{ERA5 Transformer} & \textbf{PlaSim Transformer} & \textbf{PlaSim SFNO} & \textbf{PlaSim Diffusion} \\
\hline
                    & Embedding dim $D_{emb}$      & 240 & 240 & 384 & 1024 \\
                    & Depths / blocks           & 2+6+6+2 & 2+6+6+2 & 12 & 12 self-attn.\ + 6 cross-attn. \\
Architecture        & Attention heads           & 6/12/12/6 & 6/12/12/6 & --- & 16 \\
                    & Window size               & $2\!\times\!6\!\times\!10$ & $2\!\times\!6\!\times\!12$ & --- & --- \\
                    & Patch size & $2\!\times\!2\!\times\!2$ & $2\!\times\!2\!\times\!2$ & --- & $2\!\times\!2$ \\
\hline
                    & Optimizer                 & AdamW & AdamW & AdamW & AdamW \\
Training            & Peak learning rate        & $5\!\times\!10^{-4}$ & $6\!\times\!10^{-4}$\,/\,$2\!\times\!10^{-4}$$^{\dagger}$ & $6\!\times\!10^{-4}$ & $5\!\times\!10^{-5}$ \\
                    & LR schedule               & OneCycleLR & OneCycleLR & ReduceLROnPlateau & warmup + cosine decay \\
                    & Weight decay              & 0.01 & 0.01 & 0.01 & 0.05 \\
                    & Batch size                & 32 & 32 & 32 & 32 \\
                    & Epochs                    & 120  & 120  & 200  & 800 \\
                    & Loss function             & weighted MAE& weighted MAE& spherical MSE& weighted MSE\\
\hline
Diffusion           & Steps $N_{\mathrm{diff}}$ & --- & --- & --- & 100 \\
                    & Noise schedule            & --- & --- & --- & cosine \\
\hline
\end{tabular}%
}
\vspace{2pt}
\begin{flushleft}
\end{flushleft}
\end{table}

\begin{table}[htbp]
\centering
\caption{\textbf{Architectural and training hyperparameters for the Lorenz 96 AI models.} Set notation indicates discrete values
explored during the hyperparameter optimization phase.}
\label{tab:l96_hyperparams}
\begin{tabular}{llc}
\hline
\textbf{Phase} & \textbf{Hyperparameter} & \textbf{Value / Search space} \\
\hline
              & Architecture            & multilayer perceptron, ReLU activations \\
Architecture  & Width $W$               & $\{512,1024,2048\}$ \\
              & Depth $D$               & $\{3,5,7\}$ hidden layers \\
              & Input dimension         & 8 ($X$ only) or 264 ($XY$) \\
\hline
              & Optimizer               & Adam \\
Hyperparameter search    & Learning rate           & Log-uniform $[10^{-4},10^{-2}]$ \\
              & Batch size              & $\{256,512,1024,2048,4096\}$ \\
              & Pruning                 & Hyperband (150 epochs/trial) \\
\hline
              & Max\ epochs            & 500 \\
Final training& LR schedule             & \texttt{ReduceLROnPlateau} (factor 0.5, patience 15) \\
              & Early stopping          & patience 50 epochs \\
              & Training loss           & mean squared error (MSE) \\
              & Evaluation metric       & mean absolute error (MAE) \\
\hline
\end{tabular}
\end{table}


\begin{table}[htbp]
\centering
\caption{\textbf{One-time-step RMSE of the forecast and backcast AI models.} RMSE of a single prediction step of length $\Delta t$, computed over the $X$ variables against the reference trajectory, for the selected AI models with $W{=}1024$ and $D{=}7$. Values are the mean $\pm$ standard deviation over the same 100 initial conditions as
Fig.~\ref{fig:rmse_hovmoller_1024}.}
\label{tab:single_step_rmse}
\begin{tabular}{llcc}
\hline
\textbf{$\mathbf{x}$} & \textbf{$\Delta t$} & \textbf{Forecast RMSE} &
\textbf{Backcast RMSE} \\
\hline
$X$  & $1\delta t$  & 0.0056 $\pm$ 0.0020 & 0.0056 $\pm$ 0.0021 \\
$X$  & $5\delta t$  & 0.026 $\pm$ 0.0093  & 0.027 $\pm$ 0.011 \\
$X$  & $10\delta t$ & 0.048 $\pm$ 0.016   & 0.046 $\pm$ 0.018 \\
$XY$ & $1\delta t$  & 0.038 $\pm$ 0.012   & 0.043 $\pm$ 0.015 \\
$XY$ & $5\delta t$  & 0.16 $\pm$ 0.054    & 0.23 $\pm$ 0.067 \\
$XY$ & $10\delta t$ & 0.29 $\pm$ 0.073    & 0.26 $\pm$ 0.066 \\
\hline
\end{tabular}
\end{table}

\clearpage

\begin{figure}[ht]
\centering
\includegraphics[width=\textwidth]{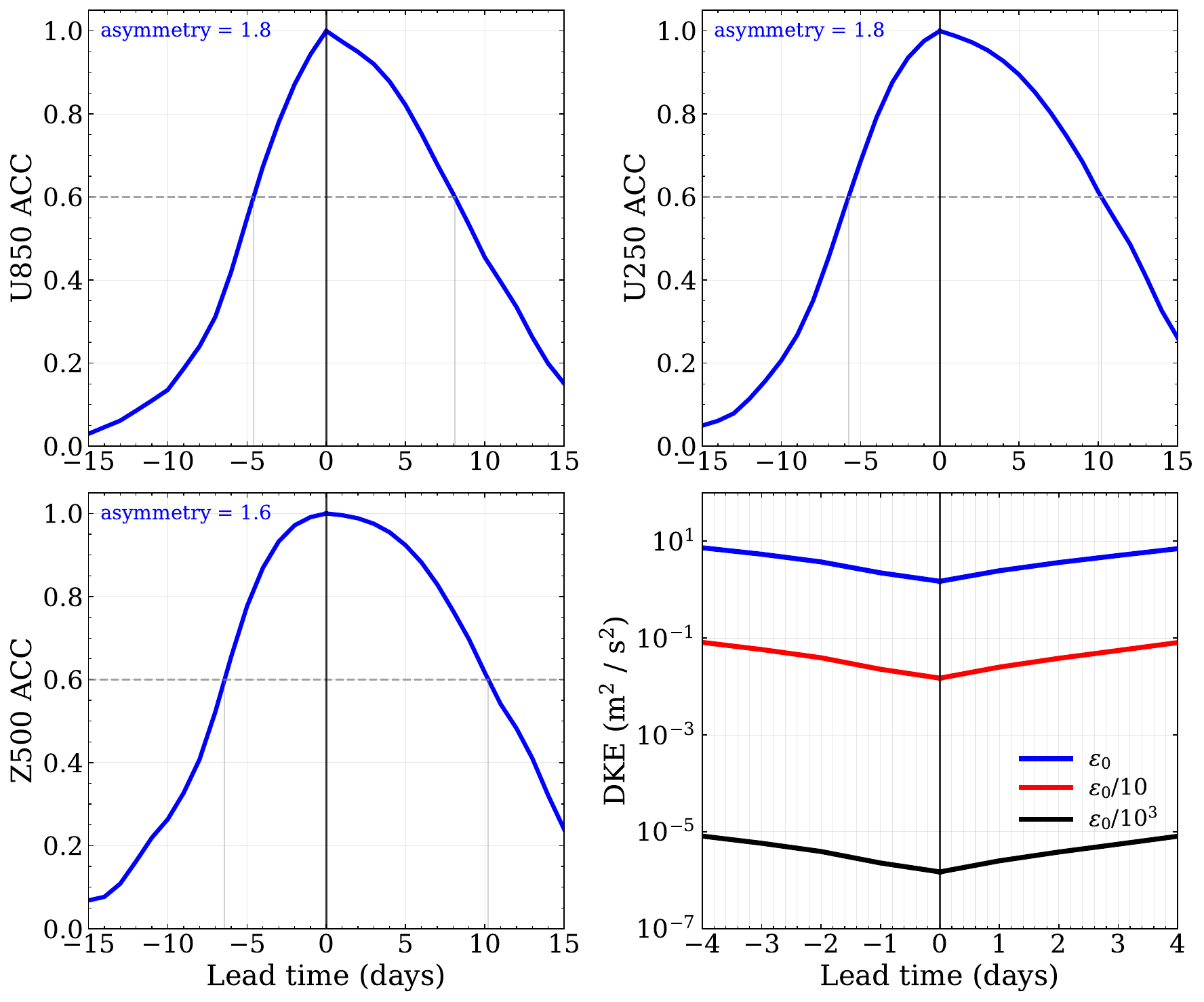}
\caption{{\textbf{Missing butterfly and forecast-backcast asymmetry in the SFNO-based AIWP models of PlaSim GCM.} As in Fig.~\ref{fig1:era5-plasim}B--C, but for independently trained forecasting and backcasting AIWP models, based on spherical Fourier neural operator (SFNO) architecture \cite{bonev2023spherical}, with $\Delta t = 24$~h (see Methods and Data). ACC of $U850$, $U250$, and $Z500$ versus lead time are shown, as well as the DKE at 300~hPa for ensembles initialized with perturbations of amplitude $\epsilon_0 = 0.1$, $\epsilon_0/10$, and $\epsilon_0/10^3$. All curves are averaged over the same initial conditions from the test set as in Fig.~\ref{fig1:era5-plasim}.}}
\label{fig:SI-SFNO}
\end{figure}
\clearpage

\begin{figure}[ht]
\centering
\includegraphics[width=\textwidth]{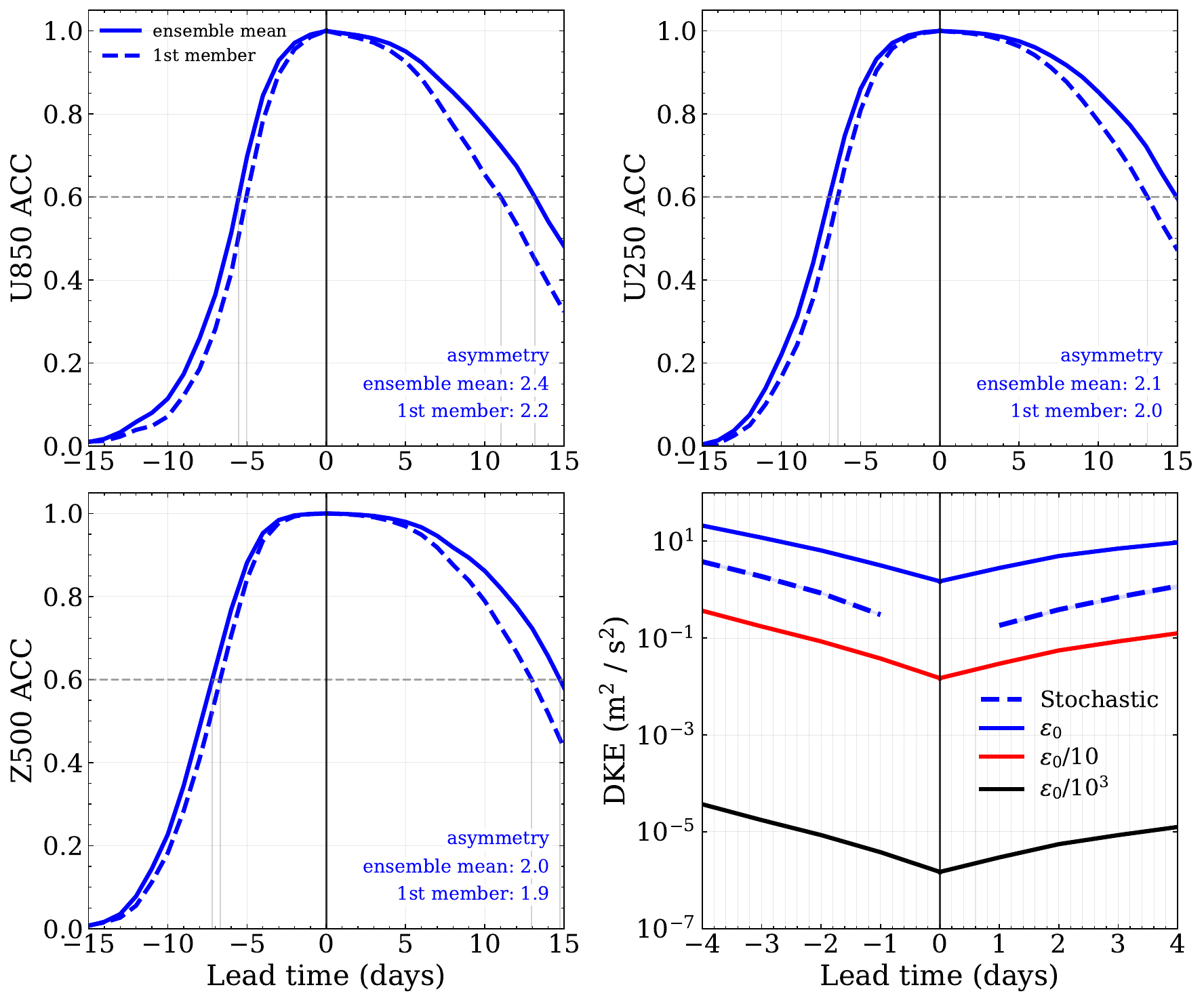}
\caption{{\textbf{Missing butterfly and forecast-backcast asymmetry in the Diffusion-based AIWP models of PlaSim GCM.} As in Fig.~\ref{fig:SI-SFNO}, but for independently trained forecasting and backcasting AIWP models, based on conditional diffusion models, with $\Delta t = 24$~h (see Methods and Data). ACC curves are shown for the 16-member ensemble mean (solid) and for a single member (dashed). In the DKE panel, solid curves are obtained in the quasi-deterministic mode used throughout the paper (see Methods and Data), whereas the dashed curve is the spread generated by stochastic sampling alone, with unperturbed initial conditions.}}
\label{fig:SI-DiT}
\end{figure}
\clearpage

\begin{figure}[ht]
\centering
\includegraphics[width=0.8\textwidth]{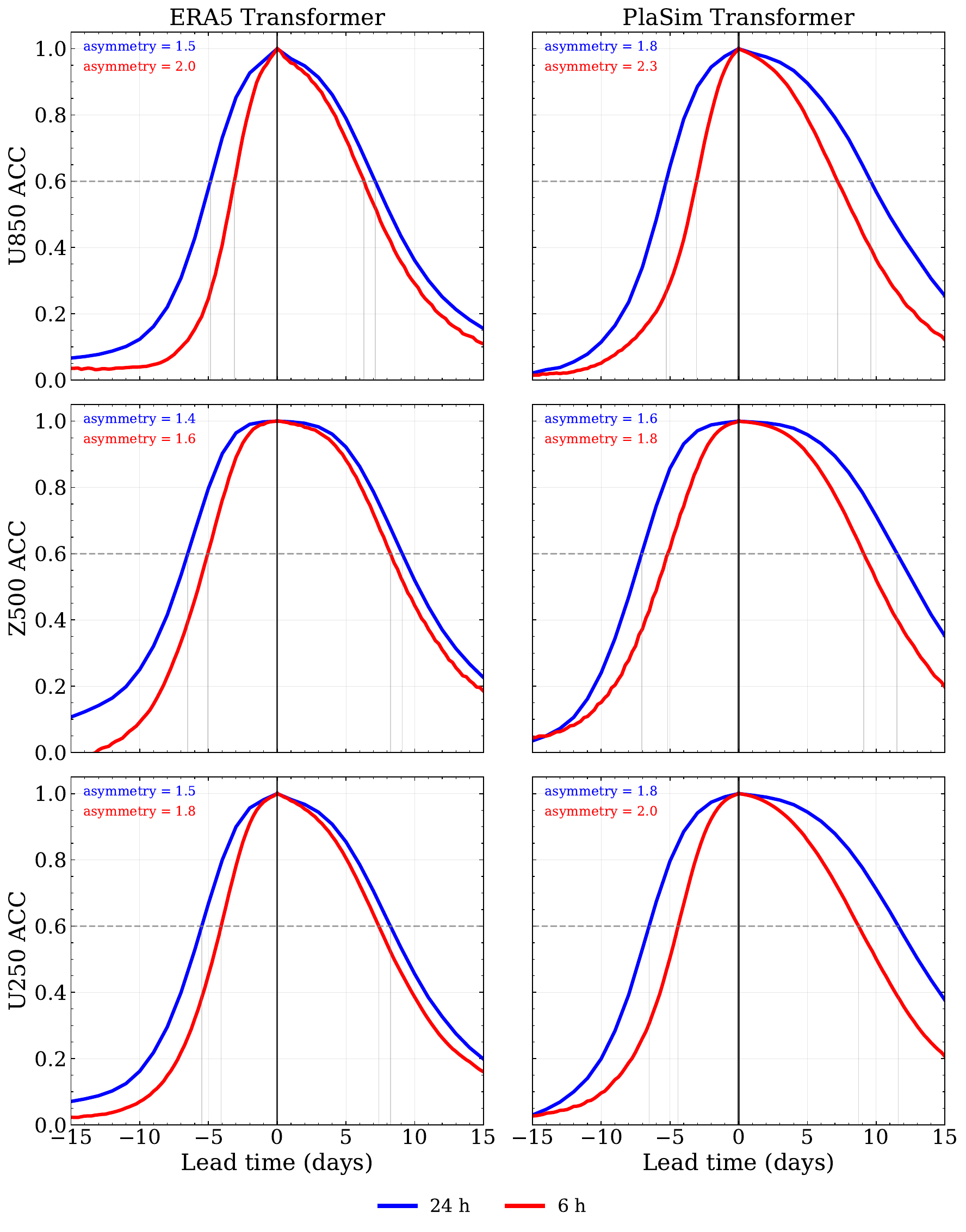}
\caption{{\textbf{Forecast-backcast asymmetry across variables and $\Delta t$ in Transformer-based AIWP models of ERA5 and PlaSim GCM.} As in Fig.~\ref{fig1:era5-plasim}B, but for U850 (top), Z500 (middle), and U250 (bottom) for ERA5 and PlaSim Transformers ($\Delta t = 24$ and 6~h). Colors denote $\Delta t$ (legend at the bottom); the color-matched annotations show the asymmetry. Curves are averaged over the test set initial conditions (same as in Fig.~\ref{fig1:era5-plasim}).}}
\label{fig:SI-ACC-asymmetry}
\end{figure}
\clearpage

\begin{figure}[ht]
\centering
\includegraphics[width=0.8\textwidth]{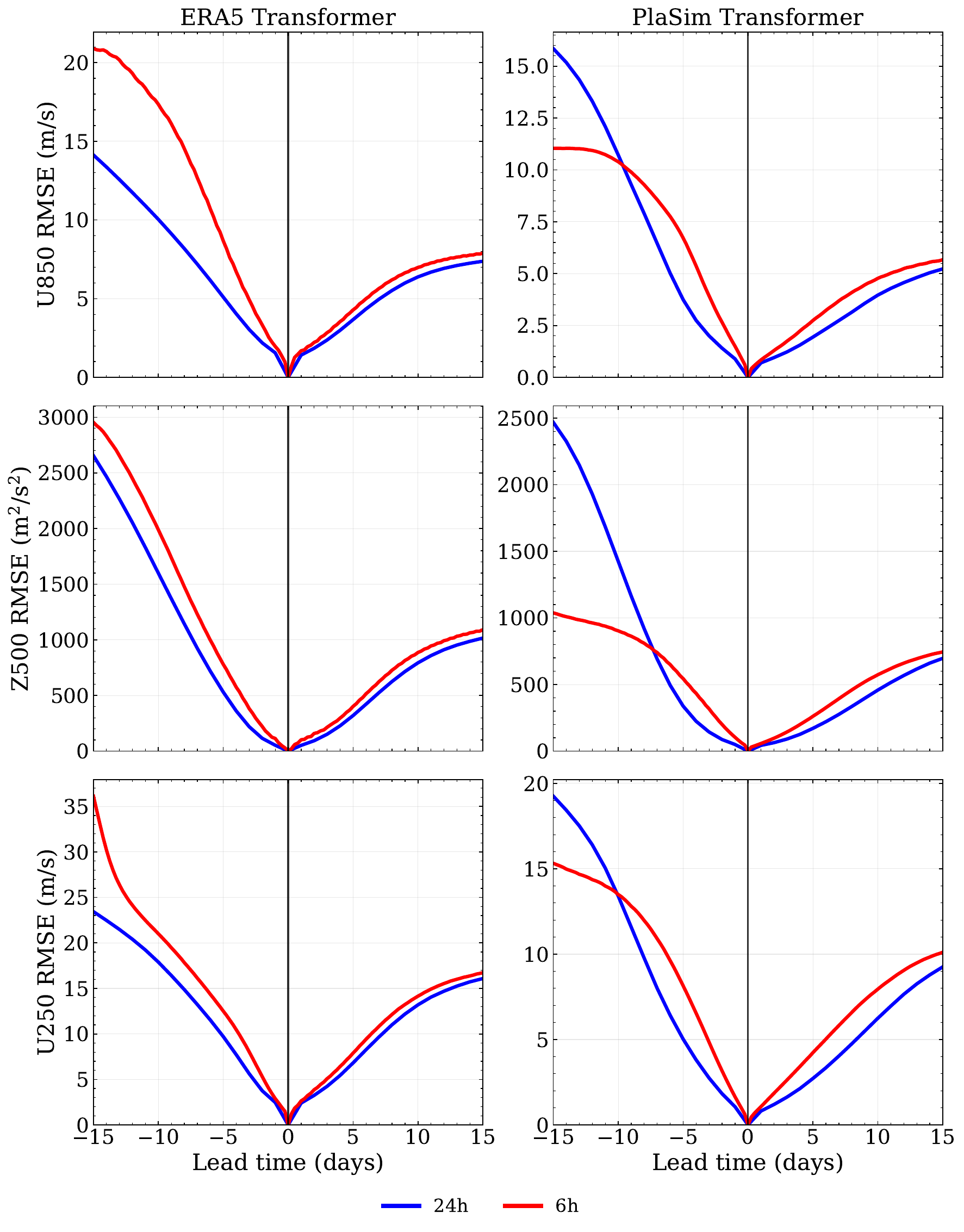}
\caption{{\textbf{RMSE-based forecast and backcast accuracy and asymmetry across variables and $\Delta t$ in Transformer-based AIWP models of ERA5 and PlaSim GCM.} As in Fig.~\ref{fig:SI-ACC-asymmetry}, but showing RMSE rather than ACC for the same models, variables, and initial conditions.}}
\label{fig:SI-RMSE-asymmetry}
\end{figure}
\clearpage

\begin{figure}[ht]
\centering
\includegraphics[width=0.65\textwidth]{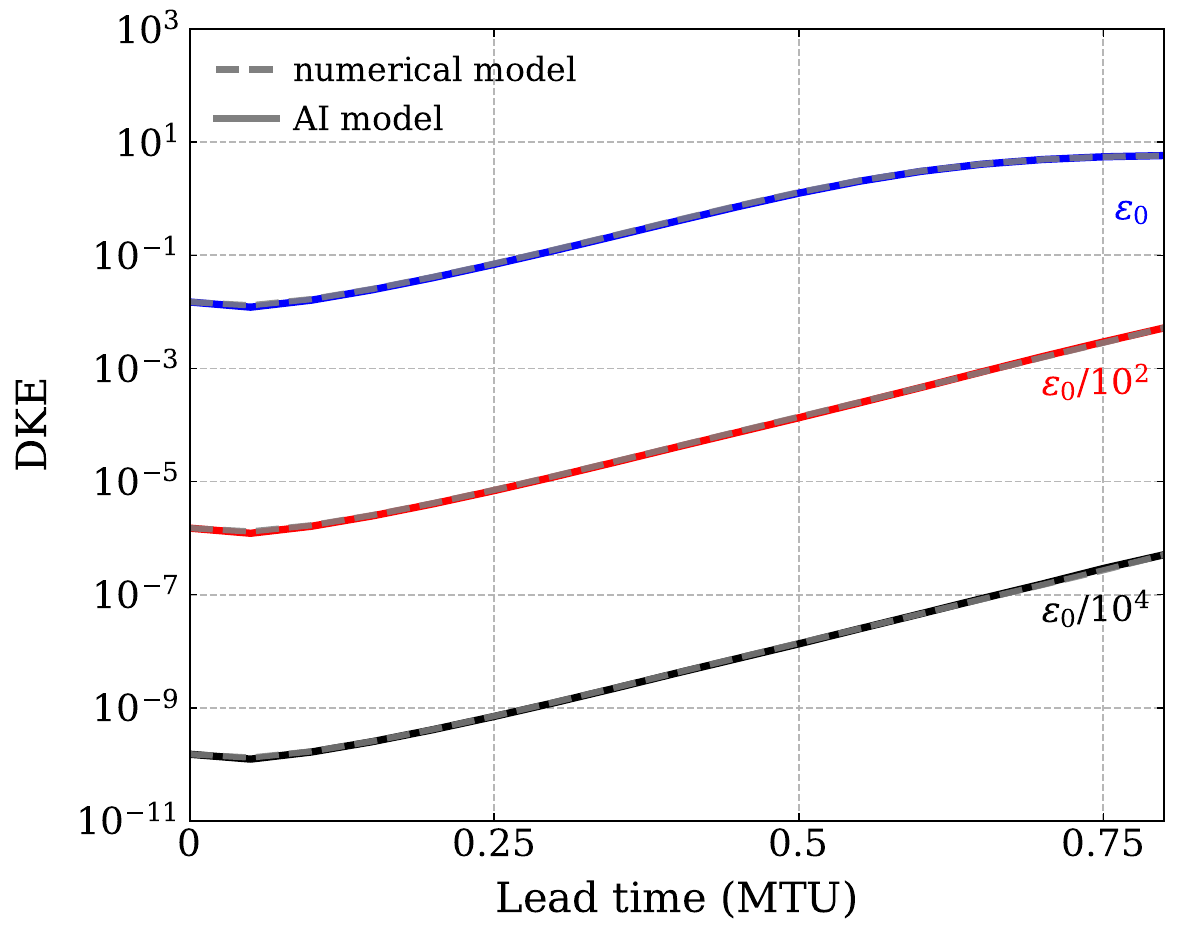}
\caption{{\textbf{DKE growth in the numerical and AI models of Lorenz 63.} Curves show DKE, computed over all three variables, of forecast ensembles for the numerical model (based on integration of Eqs.~\eqref{eq:L63x}--\eqref{eq:L63z}) and the AI model. The AI model is trained similarly to the one used for Lorenz 96 (Methods and Data): a multilayer perceptron trained on pairs of all three variables with $\Delta t = 10\delta t$. Perturbations are Gaussian, with amplitudes decreasing from a reference value, $\epsilon_0$. Curves are averaged over $10,000$ initial conditions drawn from a held-out test trajectory, each with a 100-member ensemble; the same initial conditions and perturbations are used for both models. The AI model closely tracks the numerical model's DKE curves at all amplitudes. They both lack the ``real'' butterfly effect \cite{palmer2014real, palmer2024real}.}}
\label{fig:SI-L63-DKE}
\end{figure}
\clearpage

\begin{figure}[ht]
\centering
\includegraphics[width=\textwidth]{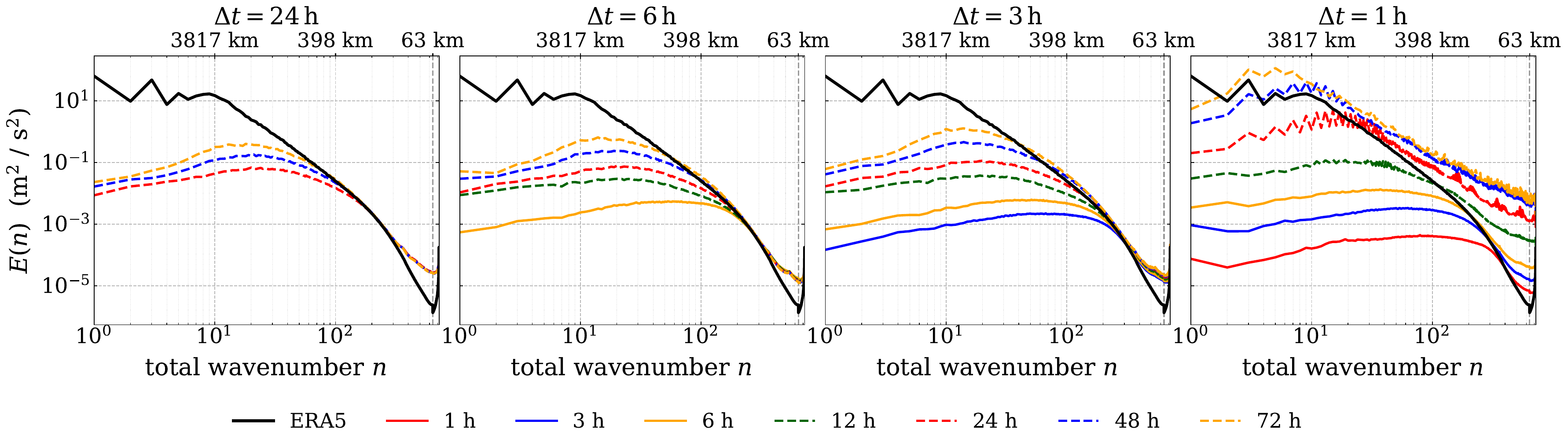}
\caption{{\textbf{Scale dependence of the forecast errors in the Pangu-Weather models with different $\Delta t$.} Power spectra $E(n)$ of the 300-hPa error field (prediction minus ERA5) as a function of total wavenumber $n$ and lead time (colors, nonuniformly from 1~h to 3~days; see legend), compared with the spectrum of the full field (black). The top axis gives the corresponding wavelength. Error saturates first at small scales and progressively fills in the larger scales; at a fixed lead time, the error is larger at all scales for the models trained with smaller $\Delta t$. Curves are averaged over the test set initial conditions.}}
\label{fig:SI-error-spectra}
\end{figure}
\clearpage

\begin{figure}[htbp]
\centering
\includegraphics[width=1.3\textwidth, angle=-90]{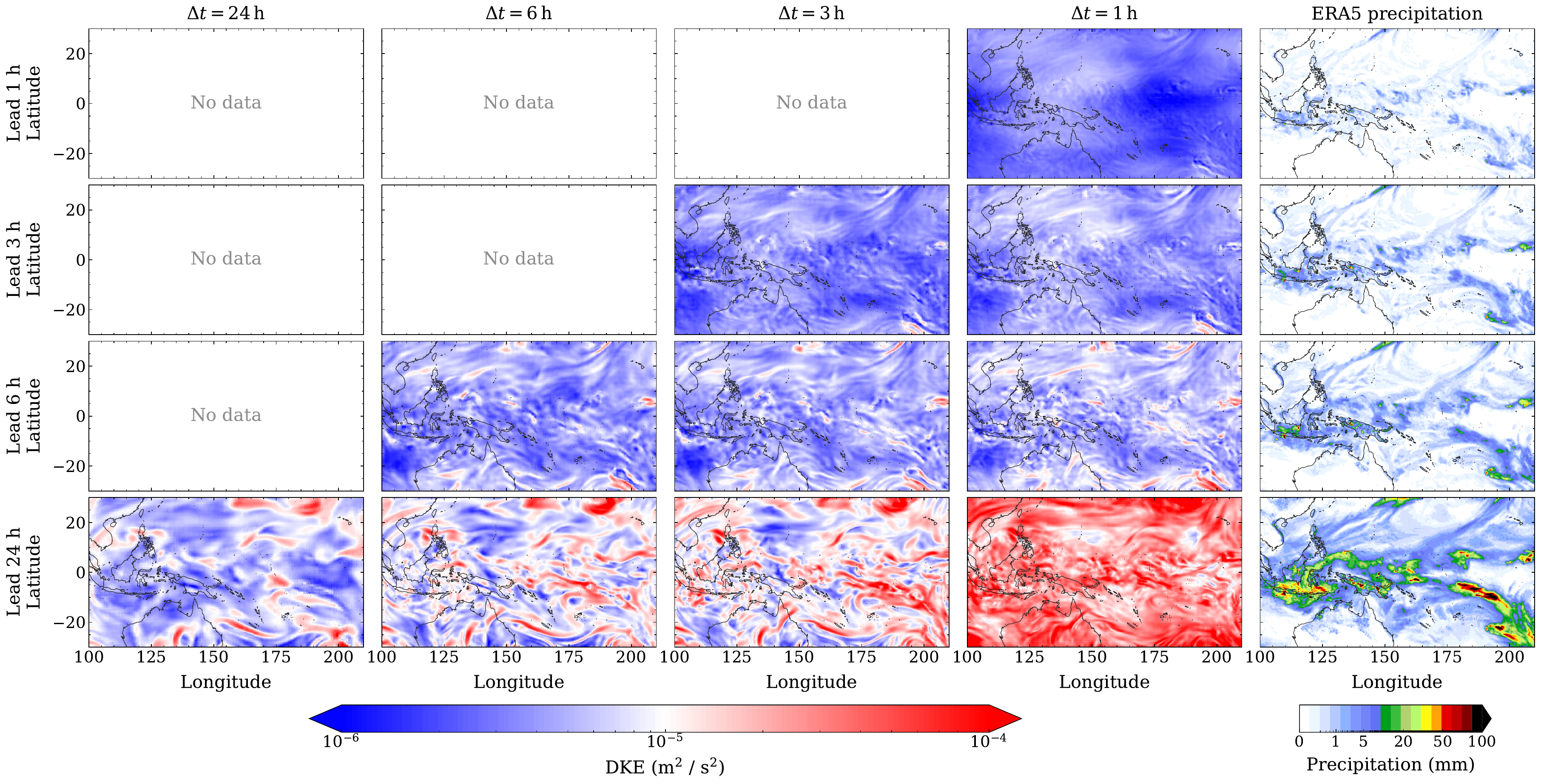}
\caption{{\textbf{Spatial distribution of Pangu-Weather forecasts' early DKE growth and the ERA5 precipitation field.} For the rotated figure, rows show lead time from 1~h (first row) to 24~h (last row). The first four columns from the left show the $\Delta t$ of the official Pangu-Weather model.  Panels marked ``No data'' correspond to lead times shorter than the model's $\Delta t$. As an example, maps of the 300-hPa DKE over parts of the tropical western Pacific are shown for ensembles with the smallest perturbation amplitude ($\epsilon_0/10^3$) for one representative initial condition. The $\Delta t = 1$~h ensembles show coherent DKE patterns rather than unphysical noise. However, unlike what numerical experiments suggest \cite{selz2023can,zhang2003effects,selz2015upscale}, the regions of largest early DKE growth in the 1-h Pangu-Weather forecasts do not necessarily collocate with precipitation. Taking the 6-h lead time as an example, the DKE pattern in the $1$-h model instead resembles a mixture of small-scale features amplified in the $6$-h model under the influence of precipitation.}}
\label{fig:SI-DKE-map+precip}
\end{figure}
\clearpage

\begin{figure}[ht]
\centering
\includegraphics[width=\textwidth]{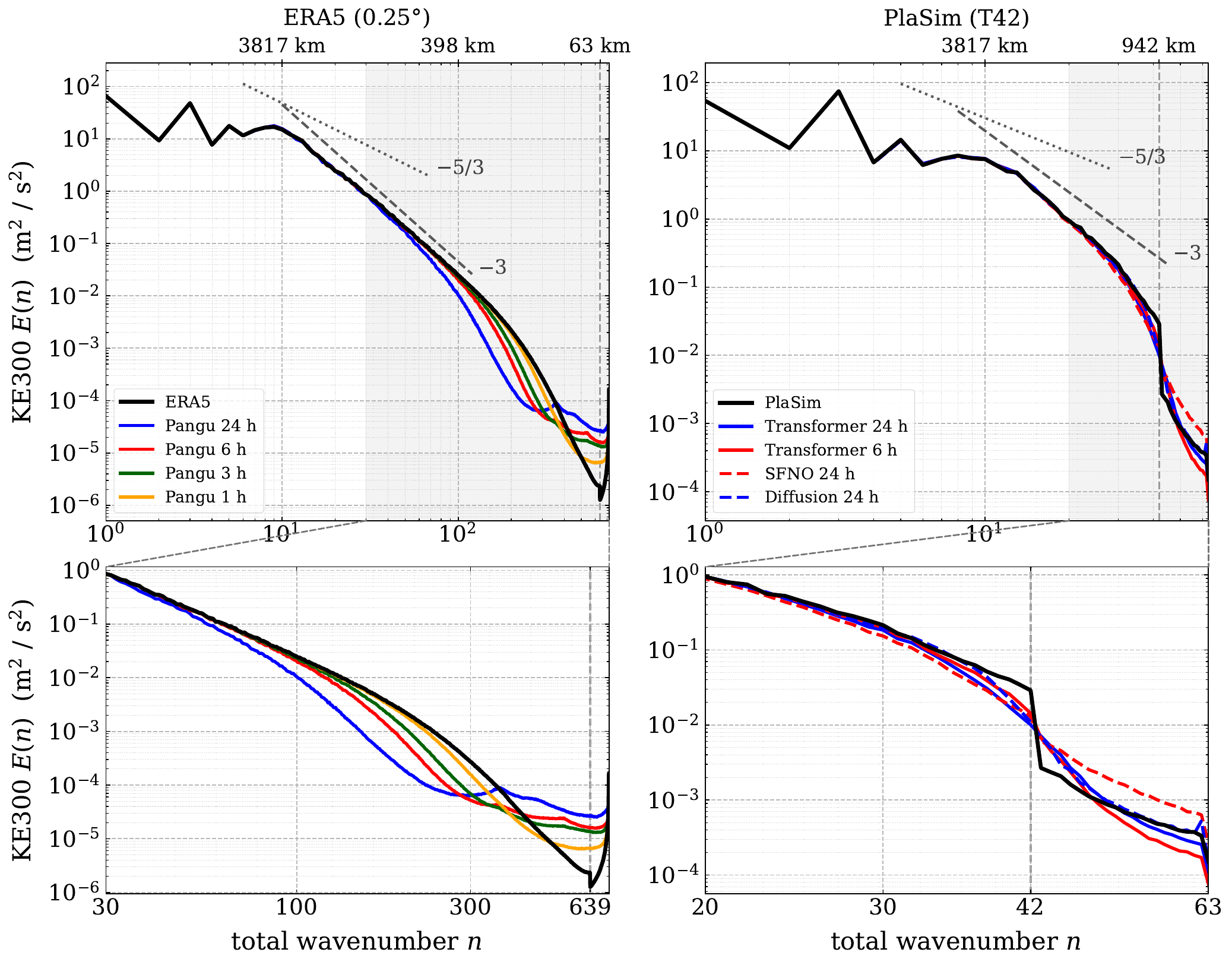}
\caption{{\textbf{Spectra of one-time-step forecasts in the AIWP models of ERA5 and PlaSim GCM.} Power spectra of the 300-hPa kinetic energy, $E(n)$, as a function of total wavenumber $n$, computed with the spherical harmonic transform, for one-time-step ($1\Delta t$) forecasts. Left: The four official Pangu-Weather models trained on ERA5 (0.25$^\circ$) from \cite{bi2023accurate}. Right: Transformer-, SFNO-, and Diffusion-based AIWP models of PlaSim GCM, which has a numerical resolution of T42. Curves are averaged
over the test set initial conditions. Bottom row zooms into the high wavenumber end for better visualization. Black curves are the reference spectra of the full ERA5 and PlaSim fields. Vertical dashed lines mark each dataset's spectral truncation (T639 for ERA5, T42 for PlaSim); gray dotted and dashed lines are $-5/3$ and $-3$ slope references.}}
\label{fig:SI-spectra-Bias-for-different-dt}
\end{figure}
\clearpage

\begin{figure}[ht]
\centering
\includegraphics[width=\textwidth]{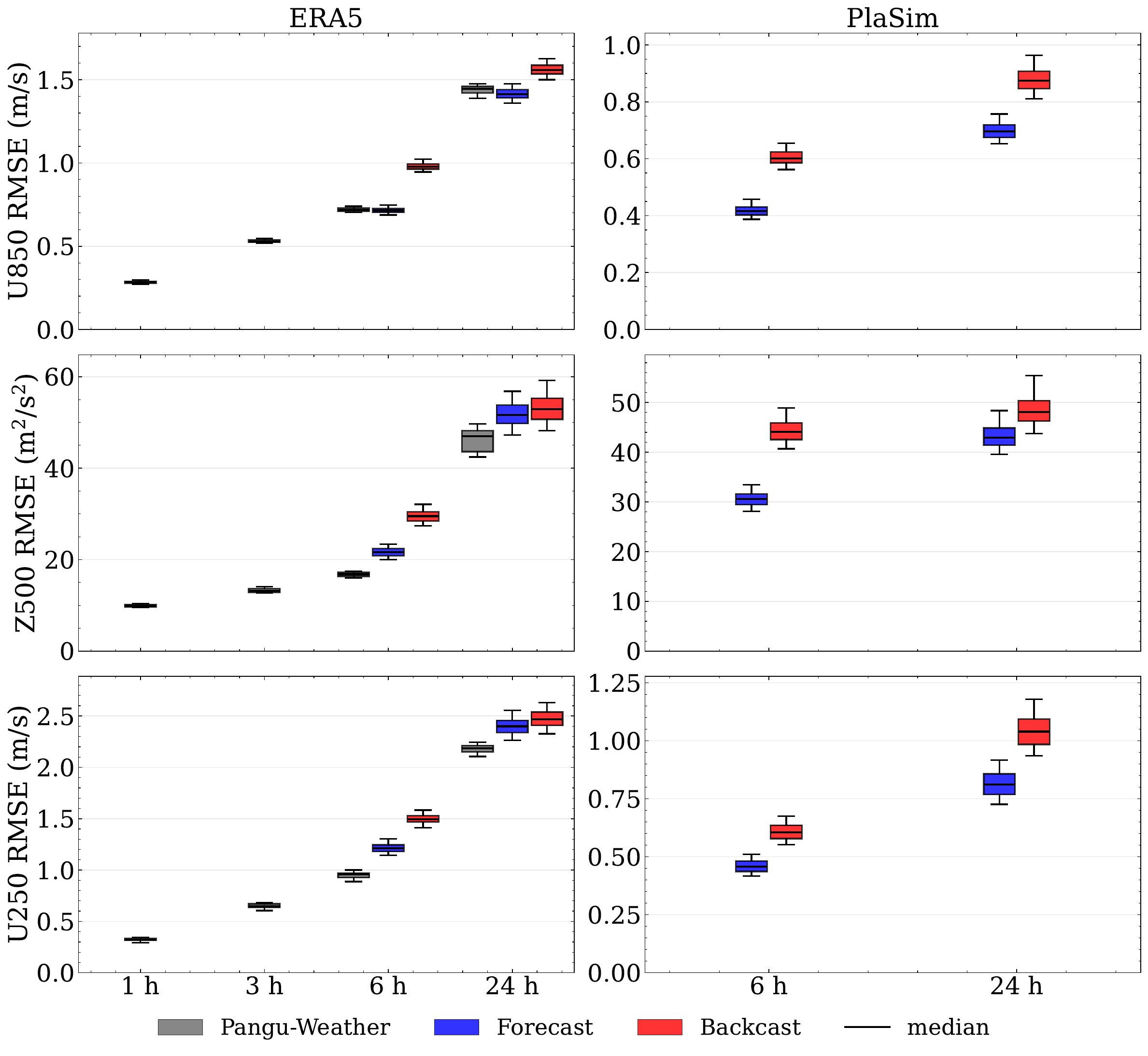}
\caption{{\textbf{One-time-step error of the forecasting and backcasting AIWP models of ERA5 and PlaSim GCM.} RMSE of one-time-step prediction ($1\Delta t$) for U850 (top),  Z500, and U250, evaluated over the test set initial conditions (see Methods and Data). Left: ERA5, which also shows forecast accuracy of the official Pangu-Weather models (gray; $\Delta t = 1$, 3, 6, and 24~h \cite{bi2023accurate}) and our forecasting (blue) and backcasting (red) Transformers ($\Delta t = 6$ and 24~h). Right: the PlaSim forecasting and backcasting Transformers ($\Delta t = 6$ and $24$~h). Boxes span the interquartile range, whiskers the 5th--95th percentiles, and the horizontal line the median.}}
\label{fig:SI-one-step-RMSE}
\end{figure}
\clearpage

\begin{figure}[ht]
\centering
\includegraphics[width=\textwidth]{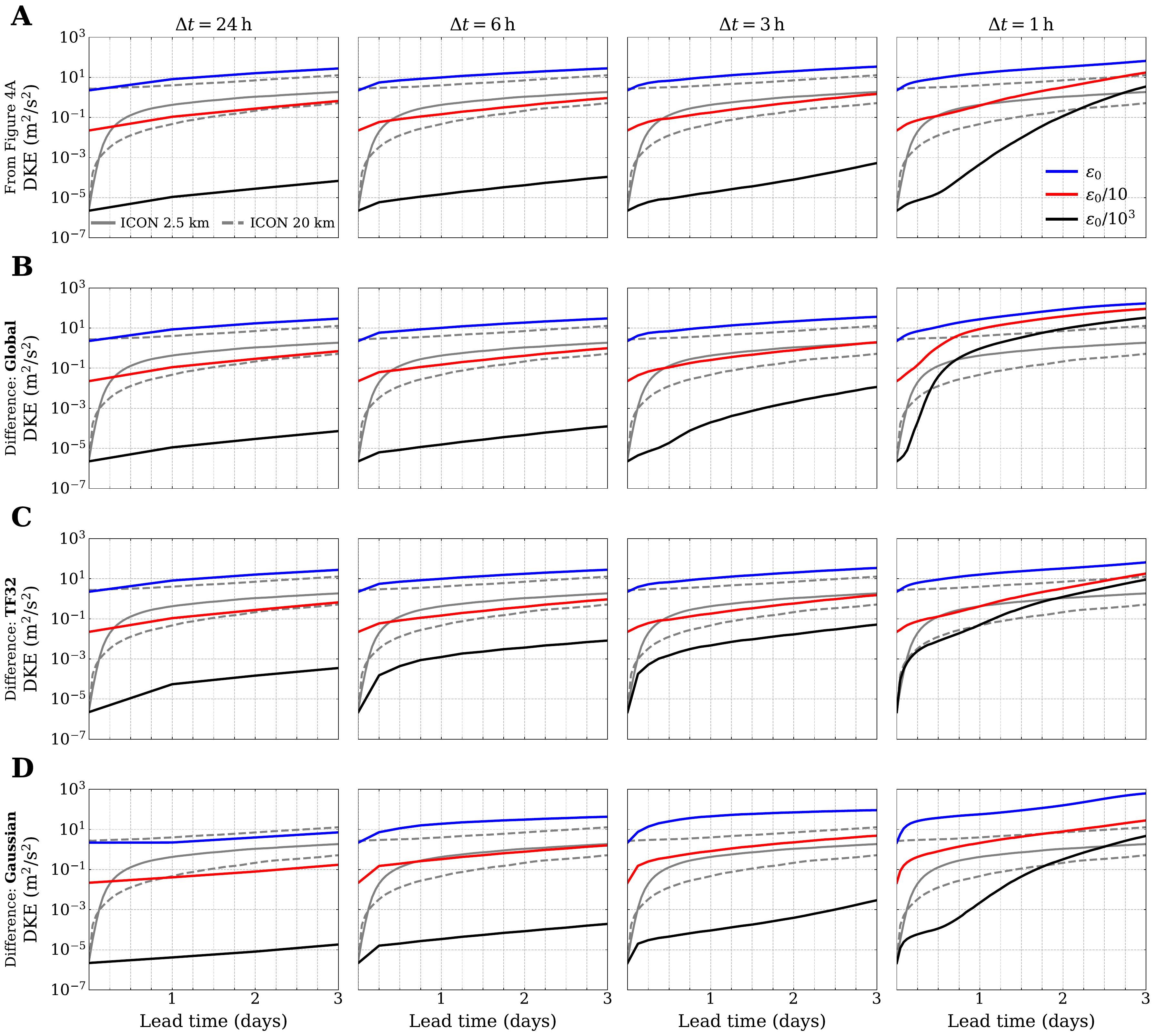}
\caption{{\textbf{Sensitivity of the Pangu-Weather forecasts' DKE growth to the averaging domain, GPU precision, and initial-perturbation structure.} Each panel shows the ensemble DKE at 300\,hPa. Columns correspond to the four official Pangu-Weather models \cite{bi2023accurate}  with $\Delta t = 24$, 6, 3, and 1\,h. \textbf{(A)}~The same as Fig.~\ref{fig:downsampling}A, the reference, which uses  DKE averaged over $60^\mathrm{o}$S-$60^\mathrm{o}$N, single precision (FP32) inference, and Perlin-noise initial perturbations. \textbf{(B)}~as in (A) but averaged globally. \textbf{(C)}~as in (A) but without strict FP32 inference, so that the GPU carries out the internal matrix operations in lower precision (TF32). \textbf{(D)}~as in (A) but with Gaussian noise instead of Perlin noise. Colors denote the initial-perturbation amplitudes. Gray curves are the ICON reference simulations, solid for 2.5~km and dashed for 20~km resolutions. Panels (B) and (C) show that the faster DKE growth for small-amplitude perturbations might be observed for the small-$\Delta t$ AIWP models, but they can be due to high-latitude unphysical instabilities or noise from lower-precision GPU calculations. (D) shows little sensitivity to the structure of the initial-condition perturbation.}}
\label{fig:SI-sensitivity-Fig4-tf32-global}
\end{figure}
\clearpage

\section*{Caption for Movie S1. Numerical forecasts and backcasts on the
Lorenz~63 attractor.}
Animation of the Lorenz~63 (Eqs.~\eqref{eq:L63x}-\eqref{eq:L63z}) states numerically integrated forward and backward from 10 initial conditions that are slightly different (same numerical methods as those used for Lorenz~96; see Methods and Data). The Lorenz attractor (thin gray lines) is visualized in the $x$--$y$--$z$ phase space. The initial conditions are on the attractor. The forecast trajectories remain on the attractor (due to the phase-space contraction), though they diverge significantly in the long term due to the $+0.91$ leading Lyapunov exponent. The backcast trajectories quickly leave the attractor, which is now a repeller (due to the phase-space expansion). See the Supplementary Materials for more discussion. The visualization code is generated by Claude Fable 5. The movie is available at \url{https://doi.org/10.5281/zenodo.22062019}.

\end{document}